\pdfoutput=1

\documentclass[prb,twocolumn,superscriptaddress,a4paper]{revtex4-2}

\usepackage{url}
\usepackage{amsmath}
\usepackage{graphicx}
\usepackage{color}
\usepackage{float}
\usepackage{soul}

\usepackage{hyperref}
\usepackage{braket}
\usepackage{siunitx}
\usepackage{bm}

\usepackage[T1]{fontenc}
\usepackage{comment}

\newcommand{\sech}[0]{\ensuremath{\operatorname{sech}}}
\newcommand{\subscript}[1]{$_{\text{#1}}$}
\newcommand{\superscript}[1]{$^{\text{#1}}$}
\newcommand{\yso}[0]{Y\subscript{2}SiO\subscript{5}}
\newcommand{\nd}[1]{\superscript{\!{#1}}Nd}
\newcommand{\yb}[1]{\superscript{\!{#1}}Yb}
\newcommand{\y}[1]{\superscript{\!{#1}}Y}

\renewcommand*{\S}[2]{\ensuremath{\mathrm{S}_{\mathrm{#1#2}}}} 
\newcommand{\D}[1]{D\subscript{#1}} 

\newcommand*{\e}{\ensuremath{\mathrm{e}}}

\usepackage{multibib}

\newcites{SM}{SM References}

\begin{document}

\title{Coherent spin dynamics of rare-earth doped crystals in the high-cooperativity regime}

\author{Joseph~Alexander}
\email[]{joseph.alexander.18@ucl.ac.uk}
\affiliation{London Centre for Nanotechnology, University College London, London WC1H 0AH, United Kingdom}

\author{Gavin~Dold}

\affiliation{London Centre for Nanotechnology, University College London, London WC1H 0AH, United Kingdom}
\affiliation{National Physical Laboratory, Hampton Road, Teddington TW11 0LW, United Kingdom}

\author{Oscar~W.~Kennedy}
\affiliation{London Centre for Nanotechnology, University College London, London WC1H 0AH, United Kingdom}

\author{Mantas~\v{S}im\.{e}nas}
\affiliation{London Centre for Nanotechnology, University College London, London WC1H 0AH, United Kingdom}

\author{James~O'Sullivan}
\affiliation{London Centre for Nanotechnology, University College London, London WC1H 0AH, United Kingdom}

\author{Christoph~W.~Zollitsch}
\affiliation{London Centre for Nanotechnology, University College London, London WC1H 0AH, United Kingdom}

\author{Sacha Welinski}
\affiliation{Universit\'e PSL, Chimie ParisTech, CNRS, Institut de Recherche de Chimie Paris, 75005 Paris, France}
\affiliation{Thales Research and Technology, 1 Avenue Augustin Fresnel, 91767 Palaiseau, France}
\author{Alban Ferrier}
\affiliation{Universit\'e PSL, Chimie ParisTech, CNRS, Institut de Recherche de Chimie Paris, 75005 Paris, France}
\affiliation{Facult\'e des Sciences et Ing\'enierie,  Sorbonne Universit\'e, 75005 Paris, France}
\author{Eloïse Lafitte-Houssat}
\affiliation{Universit\'e PSL, Chimie ParisTech, CNRS, Institut de Recherche de Chimie Paris, 75005 Paris, France}
\affiliation{Thales Research and Technology, 1 Avenue Augustin Fresnel, 91767 Palaiseau, France}

\author{Tobias Lindstr\"om}
\affiliation{National Physical Laboratory, Hampton Road, Teddington TW11 0LW, United Kingdom}
\author{Philippe Goldner}
\affiliation{Universit\'e PSL, Chimie ParisTech, CNRS, Institut de Recherche de Chimie Paris, 75005 Paris, France}

\author{John J. L. Morton}
\email[]{jjl.morton@ucl.ac.uk}
\affiliation{London Centre for Nanotechnology, University College London, London WC1H 0AH, United Kingdom}
\affiliation{Department of Electronic and Electrical Engineering, UCL, London WC1E 7JE, United Kingdom}

\date{\today}

\begin{abstract}
Rare-earth doped crystals have long coherence times and the potential to provide quantum interfaces between microwave and optical photons. Such applications benefit from a high cooperativity between the spin ensemble and a microwave cavity --- this motivates an increase in the rare earth ion concentration which in turn impacts the spin coherence lifetime. We measure spin dynamics of two rare-earth spin species, \nd{145} and \yb{} doped into \yso{}, coupled to a planar microwave resonator in the high cooperativity regime, in the temperature range \SI{1.2}{\kelvin} to \SI{14}{\milli\kelvin}. 
We identify relevant decoherence mechanisms including instantaneous diffusion arising from resonant spins and   temperature-dependent spectral diffusion from impurity electron and nuclear spins in the environment.
We explore two methods to mitigate the effects of spectral diffusion in the \yb{} system in the low-temperature limit, first, using magnetic fields of up to 1~T to suppress impurity spin dynamics and, second, using transitions with low effective g-factors to reduce sensitivity to such dynamics. Finally, we demonstrate how the `clock transition' present in the  \yb{171} system at zero field can be used to increase coherence times up to $T_{2} = \SI{6 +- 1}{\milli\second}$. 
\end{abstract}

\maketitle

\section{Introduction}

Rare-earth ions (REIs) doped into yttrium orthosilicate (\yso{}) are promising systems for use in solid-state quantum technologies. They have optical \cite{THIEL2011353} and microwave \cite{Ortu2018} spin transitions with coherence times of milliseconds for electron spin resonance transitions~\cite{Wolfowicz2015,Li2020,doi:10.1126/sciadv.abj9786}, and hours for nuclear spin transitions~\cite{Zhong2015,Rancic2017}. These properties have stimulated proposals for the use of such REIs in microwave \cite{Sangouard2011,Probst2015} or optical \cite{doi:10.1126/science.aan5959,PhysRevLett.118.210501,ruskuc_nuclear_2022} multimode quantum memories and quantum microwave to optical transducers \cite{Williamson2014,Fernandez-Gonzalvo2015,bartholomew_-chip_2020}. 
%
High efficiency storage and retrieval protocols based on spins coupled to microwave resonators require cooperativity, $C\geq 1$, and consequently high spin density~ \cite{Tavis1968, afzelius2013proposal,julsgaard2013quantum, grezes2016towards}. However, such high spin densities introduce spin decoherence mechanisms such as spectral \cite{Bottger2006} or instantaneous \cite{SALIKHOV1981255} diffusion which must be understood and mitigated to achieve useful quantum memory lifetimes.
%

In this article we perform pulsed electron spin resonance (ESR) measurements of two REI systems (\nd{145} and \yb{nat}) in YSO, using planar superconducting resonators~\cite{Dold2019} that are strongly coupled to the spin ensemble ($C\sim$ 4-245). First we measure and identify instantaneous diffusion as a limiting decoherence mechanism in highly doped spin systems. We observe a strongly temperature-dependent coherence time, and show that this is due to suppression of spectral diffusion arising from spin bath polarisation at low temperatures. Nevertheless, even at a base temperature of 14~mK, coherence times are limited by spectral diffusion from electron spin sub-ensembles. Finally, we show how so-called `clock' (or ZEFOZ - ZEro First Order Zeeman \cite{PhysRevLett.92.077601}) transitions and isotopic purification of the REI impurities can extend coherence times ($T_{2}$) at zero field to over \SI{6}{\milli\second}. 

\section{Experimental and Spin System}
\begin{figure}
	\includegraphics[width = 0.99\columnwidth]{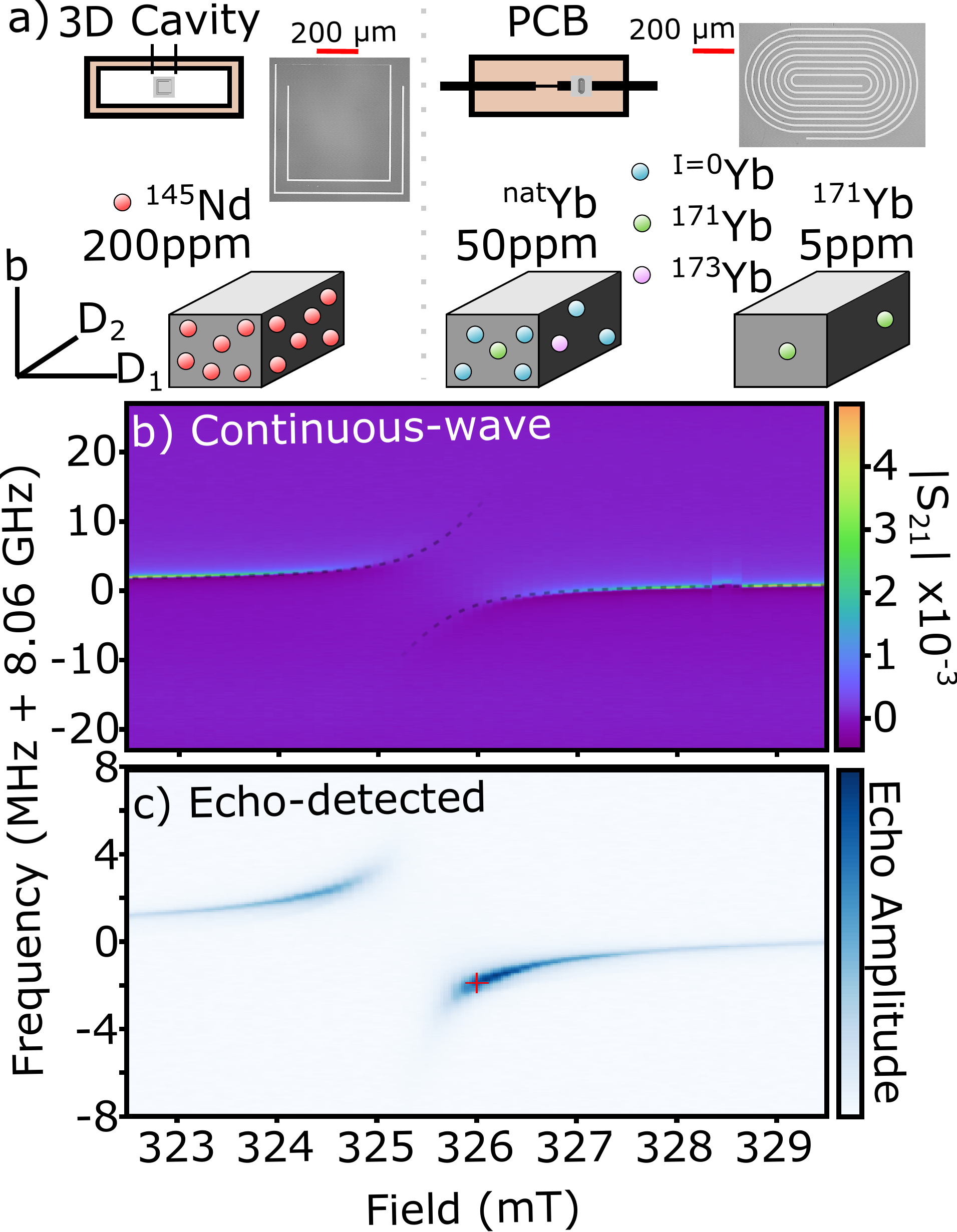}%
	\caption{a) The three rare earth ion doped YSO samples studied each had superconducting NbN resonators patterned on them to enable ESR studies of the spin systems. The \nd{145} sample (200~ppm) was measured using a `thin-ring' resonator \cite{Dold2019} and then placed inside a 3D copper cavity. The \yb{} samples were measured using a spiral resonator and placed on top of a PCB designed to maximise spin signal. b) Measurement of the \nd{145} $m_\mathrm{I} = +\tfrac{7}{2}$ transition around $B = \SI{326}{mT}$ along \D1, showing the avoided crossing in (a) continuous-wave measurement with a VNA, where the background \S21 transmission of the 3D cavity has been subtracted for clarity. c) Echo amplitude from a two-pulse Hahn echo sequence with the echo-detected avoided crossing. The darker blue regions indicate larger echo amplitude, these are points where there is more hybridisation between the resonator and the spins. The operating point for subsequent pulsed experiments on Nd is indicated by the red cross.}
	\label{fig:avcross}
\end{figure}

We study three \yso{} samples: one doped with (isotopically purified) \nd{145}\superscript{3+} ions (nuclear spin, $I=7/2$) at \SI{200}{ppm} (\SI{4e18}{\per\centi\metre\cubed}), one with natural isotopic abundance \yb{}\superscript{3+} at \SI{50}{ppm} (\SI{1e18}{\per\centi\metre\cubed}), and one with \yb{171}$^{3+}$ at \SI{5}{ppm} (\SI{1e17}{\per\centi\metre\cubed}). The crystals were cut along their principal dielectric axes ({\bf \D1}, {\bf \D2}, {\bf b}) with faces perpendicular to {\bf b} polished for resonator fabrication. Due to its larger ionic radius, \nd{145} preferentially substitutes \y{} in one of the crystal sites in \yso{}, whereas \yb{} equally populates both sites.
The \nd{} sample was studied using a `thin ring' NbN resonator~\cite{Dold2019} of film thickness \SI{45}{nm}, resonant frequency $f_\mathrm{c} = \SI{8.07}{GHz}$, and loaded quality factor $Q \approx \num{72000}$. For the \yb{} samples we used several spiral NbN resonator with film thickness \SI{15}{nm} to explore different regimes: a resonator with $f_\mathrm{c} = \SI{5.04}{GHz}$, and $Q \approx \num{31000}$, as well as lower frequency resonators targeting the zero field clock transitionwith $f_\mathrm{c} = \SI{2.372}{GHz}$, and $Q \approx \num{26000}$.
The spiral resonator design was better suited to the lower target microwave frequencies for Yb, while maintaining a footprint of about \SI{500}{\micro\metre}~$\times$~ \SI{800}{\micro\metre}.
All samples were measured using a dilution refrigerator (base temperature $\sim$14~mK) with a (3-1-1)~T vector magnet. The Nd sample was mounted on a sapphire platform inside a copper box, while the Yb samples were mounted over a copper strip line designed with impedance steps for efficient collection of signal as outlined in the Supplementary Information (SI). An overview of all three systems is shown in Fig.~\ref{fig:avcross}a).

Spin transitions were located by monitoring $S_{21}(f)$ while sweeping applied magnetic field and identifying avoided crossings between the resonator and electron spin ensemble, as shown for Nd in Fig.~\ref{fig:avcross}b). A similar measurement is possible using pulsed ESR by monitoring the two-pulse Hahn echo as a function of both magnetic field and frequency, as shown in Fig.~\ref{fig:avcross}c). The maximum echo intensity is measured at $B = \SI{326}{mT}$ where the resonator and spin line hybridise; we also note the weak background echo which can be measured far from the spin line (see SI).  Fitting this avoided crossing and extracting ensemble coupling strength ($g_{ens}$), spin linewidth ($\gamma_{s}$) and resonator linewidth ($\kappa_{c}$) returns cooperativity $C = g_\mathrm{ens}^2/\gamma_\mathrm{s}\kappa_\mathrm{c} = 245$ with $g_\mathrm{ens} > \kappa_\mathrm{c}, \gamma_\mathrm{s}$, showing the system to be in the strong coupling regime, well suited for high efficiency memories. 

%

The Nd and Yb impurities studied here are well described by the spin Hamiltonian:
\begin{equation}
    \mathcal{H} = \mathbf{S}\cdot \hat{A} \cdot \mathbf{I} +\mu_{B} \mathbf{B} \cdot \hat{g} \cdot \mathbf{S} - \mu_{n}\mathbf{B} \cdot g_{n} \cdot \mathbf{I}
\end{equation}
where the electron spin ($\mathbf{S}$) interacts with the nuclear spin ($\mathbf{I}$) described by a hyperfine tensor, $\hat{A}$, and to an external magnetic field ($\mathbf{B}$) via a g-tensor, $\hat{g}$. The nuclear spin also interacts with the field via a nuclear g-factor ($g_n$), the values of which are given in the SI. The large anisotropy of the electron g-tensor results in a ESR spectrum which has a strong angular dependence and allows for regimes in which dipole coupling between rare-earth spins can be suppressed \cite{Cruzeiro2016}. 

\section{Spin Decoherence Mechanisms}
\subsection{Instantaneous Diffusion}

\begin{figure}
    \centering
    \includegraphics[width=0.99\columnwidth]{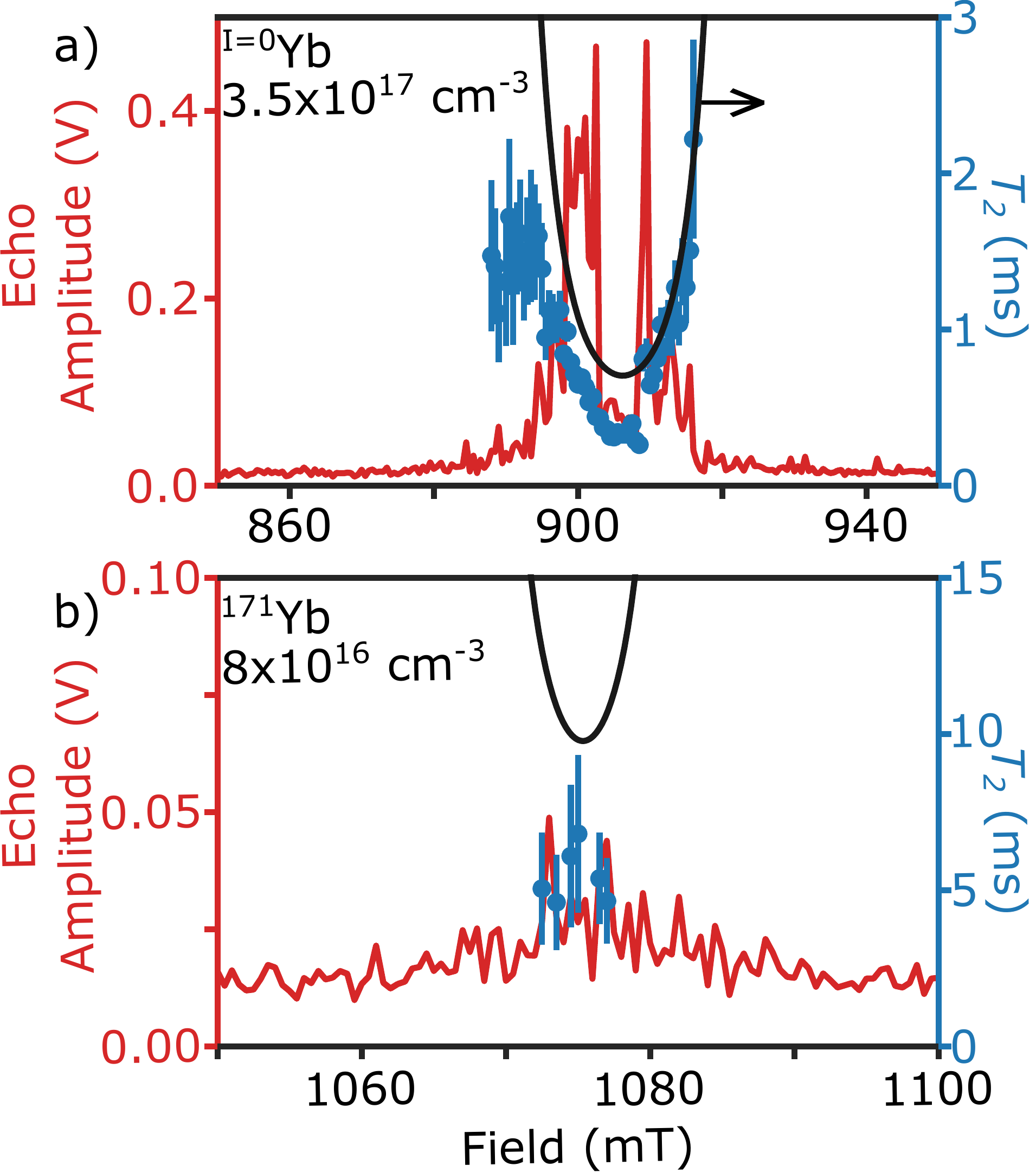}
    \caption{Echo-detected field sweep of \yb{nat} at high field (red curves). The \yb{I=0} a) and the \yb{171} b) ESR lines can be separately studied at \SI{905}{\milli\tesla} and \SI{1075}{\milli\tesla} respectively. The effect of ID can be observed as $T_{2}$ (blue datapoints) is measured across the ESR line. For \yb{I=0}, the coherence time deceases significantly at the centre of the spin line, where the density of resonant spins is maximal. In both panels, the black curve is the calculated ID limit on $T_{2}$ using Eq.~\ref{eqn:T2_ID} with no fitting parameters (see main text).}
    \label{fig:ID}
\end{figure}

A common decoherence mechanism in bulk doped spin systems --- particularly relevant in samples that are highly-doped to achieve strong coupling to resonators --- arises from the dipolar interaction between resonant spins~\cite{Schweiger2001}. Such interactions are not refocused in a two-pulse Hahn echo (because both interacting spins are flipped), leading to a decoherence mechanism known as \emph{instantaneous diffusion} (ID), whose rate is proportional to $n$, the density of resonant spins, as~\cite{Schweiger2001,Tyryshkin2011a}:
\begin{align}\label{eqn:T2_ID}
    T_{2} = \frac{9\sqrt{3}\hbar}{\pi\mu_{0}(g\mu_{\mathrm{B}})^{2}n}.
\end{align}
Here, $g$ is the effective g-factor, while $h$, $\mu_{0}$ and $\mu_{\mathrm{B}}$ are Planck's constant, the vacuum permeability and Bohr magneton, respectively.

We use two approaches to study the effect of resonant spin density on coherence time: i) exploiting differences in the natural abundance of those \yb{} isotopes which can be spectrally separated due to their nuclear spin; and ii) using the variation of spin density across the inhomogenously broadened ESR transition. The resonant spin density can be estimated based on the product of the impurity concentration of Yb, the natural abundance of the relevant isotope(s), and the fraction of spins which are addressed given a particular excitation/resonator bandwidth. For example, here a bandwidth of about \SI{66}{\kilo\hertz} is set by the microwave $\pi$-pulse length of \SI{15}{\micro\second}, which is to be compared against the inhomogenously broadened spin linewidth of $\gamma_{s}/2\pi$ = \SI{8.7}{\mega\hertz}.

In Fig.~\ref{fig:ID} we show echo-detected magnetic field sweeps across two ESR transitions: one arising from the set of Yb isotopes with nuclear spin $I=0$
(17.5~ppm) and the other from \yb{171} (3.5~ppm). 
The resonator frequency shifts as the spin transition is approached (as illustrated for Nd in Fig.~\ref{fig:avcross}(c), so the applied microwave frequency is adjusted at each magnetic field value to follow the resonator. The drop in echo amplitude seen in the centre of the line in Fig.~\ref{fig:ID}(a) arises from the strong resonator/spin hybridisation and not a result of a distortion of the ESR lineshape, which we assume here to follow a simple Gaussian profile.  
%
We  measure $T_{2}$ across the inhomogenously broadened ESR lineshape and observe a drop in the coherence time at the centre, where the resonant spin density is greatest. Indeed, using the nominal \SI{50}{ppm} concentration of \yb{}\superscript{3+}, natural isotopic abundance, expected excitation bandwidth, and measured ESR linewidth with Eq.~\ref{eqn:T2_ID} we can accurately reproduce the observed $T_{2}$ values with no free parameters. Towards the tails of the ESR line, where the resonant spin density is lowest, the measured $T_{2}$ begins to deviate from that limited by ID and saturates at 1--2~ms.
%
Performing the same measurement for the \yb{171} transition we observe no change in $T_{2}$ over the ESR lineshape --- this is expected because we can use the results above to predict an ID-limited $T_{2}$ of \SI{11}{\milli\second} (thanks to the lower spin concentration of \yb{171}), which is well above the measured values of \SI{6 +- 2}{\milli\second}. We therefore confirm that ID is not a limiting decoherence mechanism in the \yb{171} system for this sample and move on to explore alternative mechanisms.

\subsection{Spectral diffusion}
\subsubsection{Temperature dependence}

\begin{figure*}
	\includegraphics[width = 0.8\textwidth]{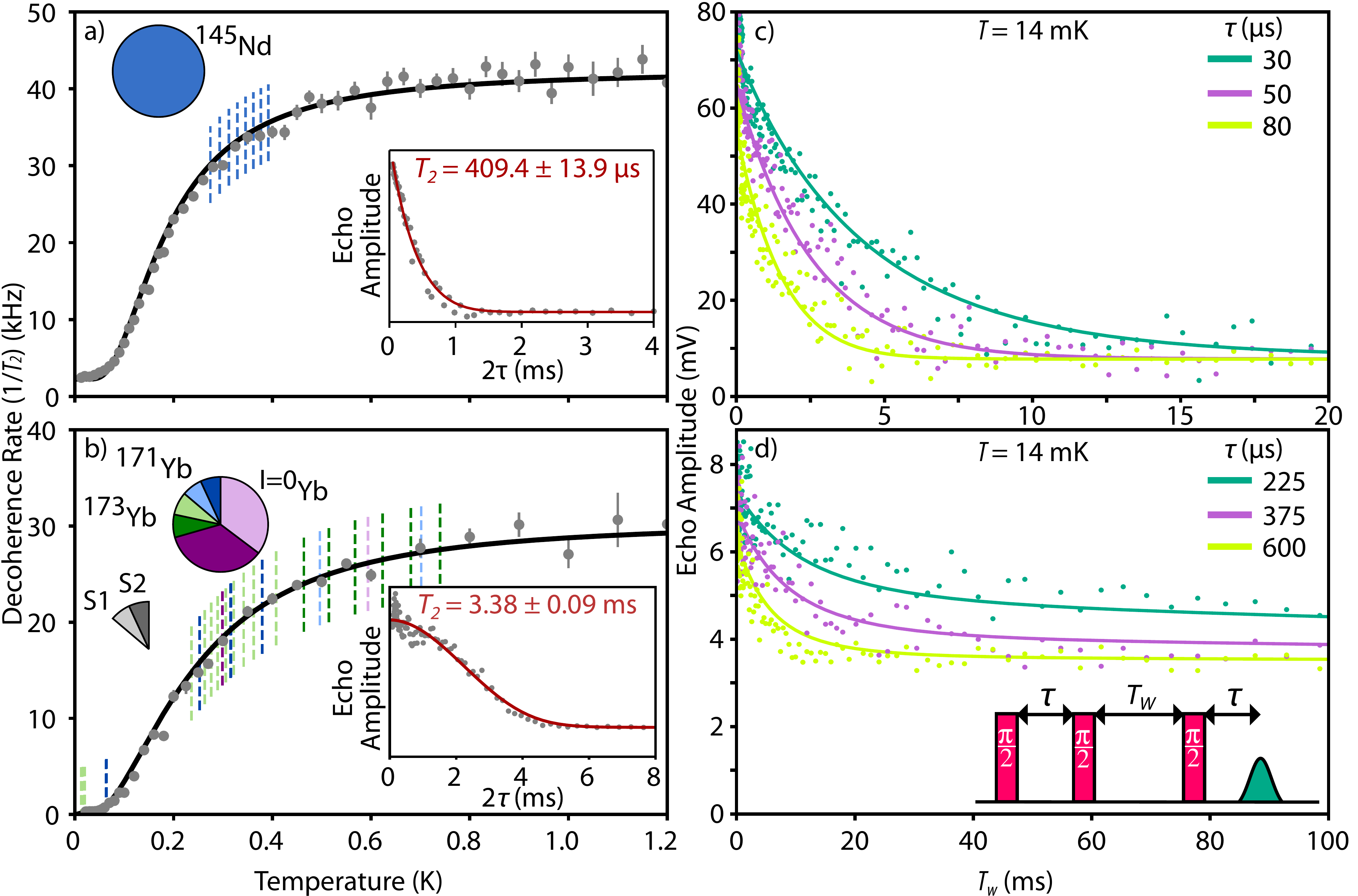}%
	\caption{The decoherence rate ($1/T_2(T)$) measured from \SI{1.2}{K} to \SI{14}{mK} for (a) the \nd{145} sample and (b) the \yb{nat} sample. In both samples, the decoherence rate drops with temperature due to suppression of spectral diffusion below the Zeeman temperature of various spin sub-ensembles present in the sample. The solid curve is a fit using Eqn.~\ref{eqn:t2oft}, and dashed lines indicate the Zeeman temperatures of all the sub-ensembles used in the fit, whose relative abundance is shown in pie charts. The inset on a) and b) show the measurement of $T_2$ at the lowest temperature \SI{14}{mK} using a two-pulse echo sequence. (c) and (d) show stimulated (3-pulse) echo decays for the \nd{145} and \yb{nat} samples, respectively, aiming to determine the source of decoherence at the base temperature of 14~mK. The values of $\tau$ in the 3-pulse echo sequence were chosen for each sample in proportion with $T_{2}$  at 14~mK. For \yb{nat} (\nd{145}), the magnetic field was 370~mT (326~mT) and microwave frequency 5.04 (8.07) GHz.
	}
	\label{fig:T2T}
\end{figure*}

Cross-relaxation processes (or spin flip-flops) between spins cause the local magnetic environment to fluctuate over time giving rise to \emph{spectral diffusion} \cite{Schweiger2001,Bottger2006,Lim2017}. Reducing the temperature of the spin bath polarizes it, reducing the rate of flip-flops and consequently, the effects of spectral diffusion \cite{Takahashi2008,rancic_electron-spin_2022}.
Figure~\ref{fig:T2T} shows such an increase in coherence time with decreasing temperature, for both the \nd{145} and \yb{nat} samples: in Nd from \SI{24(1)}{\micro\second} at \SI{1.2}{K} to \SI{0.41(1)}{\milli\second} at \SI{14}{mK} and in \yb{nat} increasing from \SI{33(7)}{\micro\second} to \SI{3.4(1)}{\milli\second} over the same temperature range. 
As in Ref.~[\citenum{Probst2015}], we model the temperature dependence of $T_2$ assuming that spectral diffusion is caused by one or more sub-ensembles of spins with distinct resonant frequencies. 
In our $^{145}$Nd sample, the different projections of the $I=7/2$ nuclear spin along the applied magnetic field form the dominant contribution to such sub-ensembles, while in the \yb{nat} sample, multiple isotopes with different nuclear spin and the projections of nuclear spin for $I>0$ act as sub-ensembles. 

We can write the decoherence rate $\Gamma=1/T_2$ of the `central spin' being studied as~\cite{Bottger2006,Lim2017}:
\begin{equation}
    \Gamma = \frac{R\Gamma_{\mathrm{SD}}}{2} \left ( \sqrt{\Gamma_{0}^{2} + \frac{R\Gamma_{\mathrm{SD}}}{\pi}} - \Gamma_{0} \right )^{-1} \approx \frac{\sqrt{\pi R \Gamma_{\rm SD}}}{2},
    \label{eqn:T2SD}
\end{equation}
where $R$ is the spin flip rate, $\Gamma_{\rm SD}$ is the spectral diffusion linewidth, and $\Gamma_0$ is a residual decoherence rate in the absence of spectral diffusion. When the dominant noise process is spectral diffusion (i.e. $\Gamma_0 \ll \sqrt{R\Gamma_{\rm SD}}$), we neglect the $\Gamma_0$ dependence of $T_2$.
Assuming that spectral diffusion occurs from spin flips due to spin-spin interactions between resonant sub-ensembles, we can write the flip-flop rate within the $i$th sub-ensemble of spins as \cite{Cruzeiro2016}:
\begin{equation}
    R_i = \beta_{\rm ff}\left ( \theta \right ) \frac{n_i^2}{\Gamma_i}\sech^2\Big(\frac{T_{\mathrm{Z},i}}{T}\Big)
    \label{eqn:R}
\end{equation}
where $n_i$ and $\Gamma_i$ are, respectively, the spin density and linewidth of the $i$th sub-ensemble, and $\beta_{\rm ff}$  is a parameter which encompasses the angular dependence of the coupling parameter and the g-tensor. $T_\mathrm{Z,i}$ is the Zeeman temperature for each sub-ensemble - $T_{z,i} = hf/k_{B}$. In the case of an isotropic medium $\beta_{\rm ff} \propto \mu_{B}^{4}g^{4}$, however, this does not hold in the case of strong anisotropy, as in YSO \cite{Sun2008,Wolfowicz2015,Welinski2016,PhysRevB.100.165107}. This term is proportional to the rate derived from Fermi's golden rule, taking the dipole coupling as a perturbation of the spin Hamiltonian. To account for this we rewrite $\beta_{ff} = \xi\frac{\mu_{B}^{4}\mu_{0}^{2}}{h^{2}}M_{i}^{2}$ where $M_{i}$ is the matrix element of each sub-ensemble transition, 
and $\xi$ is a coupling rate fit parameter. We thus rewrite $R$ as: 

\begin{equation}
    R_{i} = \xi \frac{\mu_{B}^{4}\mu_{0}^{2}}{h^{2}} M_{i}^{2} \frac{n_{i}^{2}}{\Gamma_{i}} \sech^2\Big(\frac{T_{\mathrm{Z},i}}{T}\Big).
\end{equation}

The spectral diffusion linewidth of the $i$th sub-ensemble (the frequency shift of the central spin by the spins in this sub-ensemble due to dipolar interactions) is given by~\cite{Bottger2006}:
\begin{equation}
    \Gamma_{\mathrm{SD},i} = \frac{\pi\mu_0\mu_B^2}{9\sqrt{3}h}n_ig_ig\sech^2\Big(\frac{T_{\mathrm{Z},i}}{T}\Big),
    \label{eqn:gammaSD}
\end{equation}
where $g$ and $g_i$ are the effective g-factors of the central and $i$-th spins, respectively. Combining Eqs.~\ref{eqn:T2SD}--\ref{eqn:gammaSD} gives an expression for the overall temperature-dependent decoherence rate, where the spectral diffusion component contains contributions from multiple sub-ensembles, similar in form to that given in Ref.~\cite{Probst2015}:
\begin{equation}\label{eqn:t2oft}
\Gamma(T) = \Gamma_\mathrm{res} + \sqrt{\frac{\pi\mu_0^{3}\mu_B^6}{9\sqrt{3}h^{3}\Gamma}g}\sum_i \frac{\sqrt{\xi} n_i^{3/2}M_i}{\Big(1+\e^{\tfrac{T_\mathrm{Z}^{i}}{T}}\Big)\Big(1+\e^{-\tfrac{T_\mathrm{Z}^{i}}{T}}\Big)}.
\end{equation}
$\Gamma_\mathrm{res}$, which we treat as a fit parameter, is a temperature-independent decoherence rate due to, for example, spectral diffusion from spins that remain unpolarised even at our base temperature of 14~mK. $\xi$ is a single temperature-independent fit parameter reflecting the average effective g-factor ($g_{i}$) of all sub-ensembles. $M_{i}$ was calculated using EasySpin~\cite{Stoll2006}. We assume the linewidth of each sub-ensembles is equal to that of the central spin.
The relative populations of the ground $P_\downarrow$ and excited $P_\uparrow$ states involved in the flip-flop processes for each sub-ensemble are $P_\downarrow = [1 + \exp(+T_\mathrm{Z}^{i}/T)]^{-1}$ and $P_\uparrow = [1 + \exp(-T_\mathrm{Z}^{i}/T)]^{-1}$ according to Boltzmann statistics \cite{Takahashi2008,Probst2015}. 

Fits to Eqn.~\ref{eqn:t2oft}, where $\xi$ and $\Gamma_{\rm res}$ are the only free parameters, are shown in Fig.~\ref{fig:T2T}. The resulting free parameter $\xi$ returns values of
\SI{1.94+-0.01}{} for \nd{} and \SI{12+-1}{} for \yb{}. As $\xi \propto g^{4}$, the difference in these values is attributed to the differences in the g-tensor components, in particular $g_{z} = 4.17$ in \nd{} whereas $g_{z} = 6.06$ in \yb{} site 2 ~\cite{Wolfowicz2015,Welinski2016}.
Notably, the Nd sample also has substantially larger residual decoherence rate, $\Gamma_\mathrm{res}$, at the base temperature of the dilution fridge than the Yb sample, however, we believe this is partially due to poor thermalisation in the \nd{} sample, as outlined in the SI.

\subsubsection{Low-temperature limit}

\begin{figure}
	\includegraphics[width = 0.99\columnwidth]{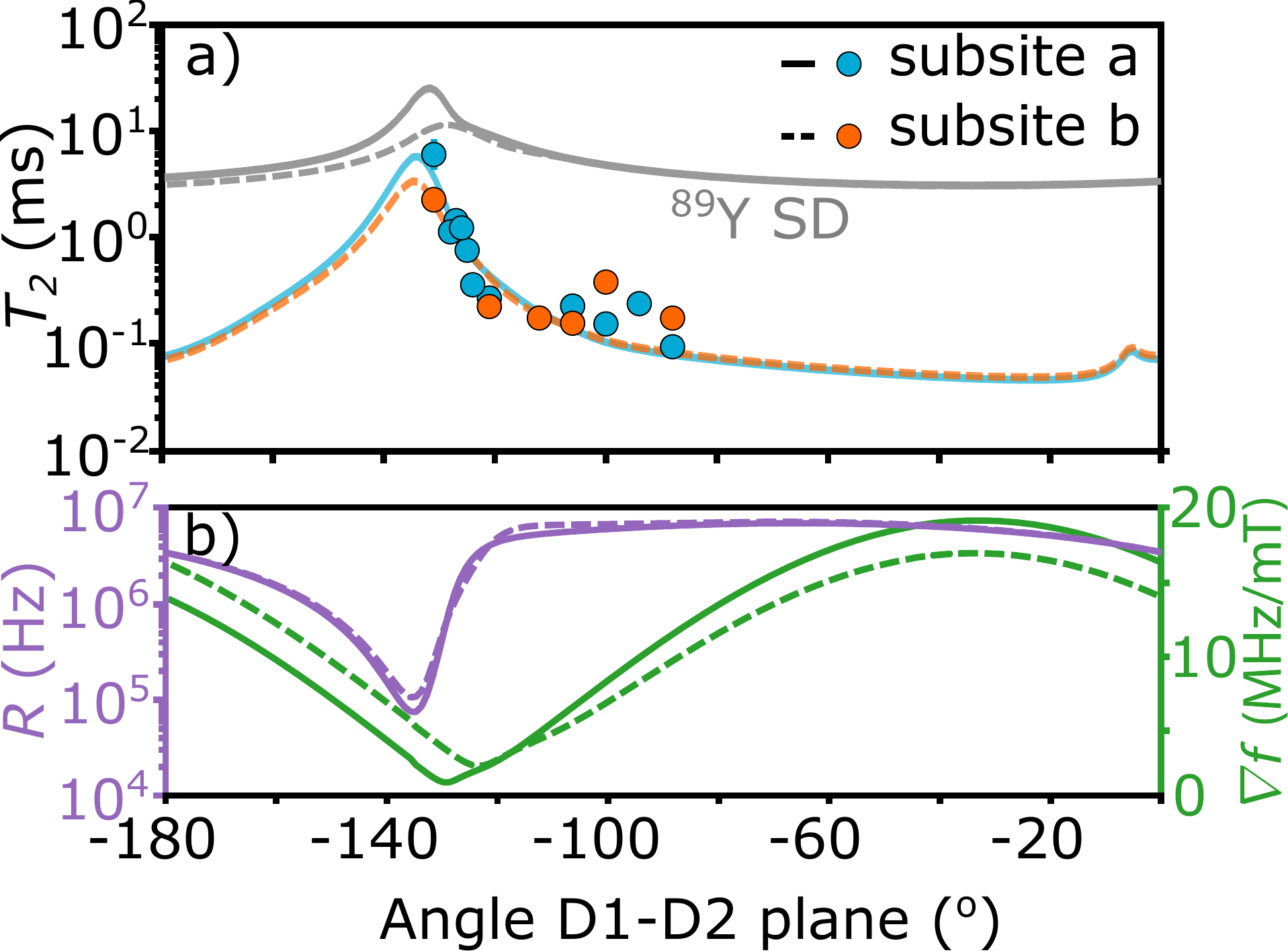}%
	\caption{Angular dependence of $T_2$ in the high field regime. a) The measured coherence time of ~\yb{171} as the applied field is rotated in the D1-D2 plane, the subsite degeneracy is lifted due to a slight misalignment with the b axis. At \SI{-131}{^{\circ}}, $T_{2}$ reaches a maximum, which corresponds well with the modelled coherence time extracted using $R\Gamma_{SD}$. b) As the angle of the applied field is rotated in the D1-D2 crystal plane, the anisotropic g-tensor results in an angular dependence of the spectral diffusion rate ($R$) and the `\yb{171}'s sensitivity to it ($\nabla f$). At angles of about \SI{131}{^{\circ}}, both $R$ and $\nabla f$ reach a minima resulting in an increase in the coherence time.}
	
	\label{fig:fig3-angular_dep}
\end{figure}

To further investigate the residual decoherence mechanism present at low temperatures, we used a three-pulse stimulated echo sequence ($\tfrac{\pi}{2}$--$\tau$--$\tfrac{\pi}{2}$--$T_\mathrm{w}$--$\tfrac{\pi}{2}$--$\tau$--echo) \cite{Lim2017} to directly measure the component due to spectral diffusion. The inhomogeneous fields induced by planar resonators on bulk-doped crystals means that the pulses are not true $\tfrac{\pi}{2}$ pulses. The stimulated echo amplitude is given by \cite{Lim2017}:
\begin{align}\label{eqn:3pulse}
A(\tau,T_\mathrm{w}) &= A_0 \exp \left[ -\left( \frac{T_\mathrm{w}}{T_1} + 2\pi \tau \Gamma_\mathrm{eff} \right) \right],
\end{align}
where $A_0$ is a fitted amplitude, $\tau$ and $T_\mathrm{w}$ are delay parameters used in the pulse sequence, and $T_1$ is the spin relaxation time measured to be \SI{696}{\milli\second} (\nd{}) and \SI{47}{\milli\second} (\yb{}) from data shown in the SI. The effective decoherence rate, $\Gamma_\mathrm{eff}$, is:
\begin{align}\label{eqn:gamma_eff}
\Gamma_\mathrm{eff} &= \Gamma_0 + \tfrac{1}{2}\Gamma_\mathrm{SD} (R\tau + 1 - \e^{-RT_\mathrm{w}}),
\end{align}
where $\Gamma_\mathrm{SD}$ is the spectral diffusion linewidth, $R$ is the total spin-flip rate, and $\Gamma_0$ is a residual decoherence rate. We assume all decoherence occurs from spectral diffusion and set $\Gamma_{0}=0$.

As shown in the SI, fitting Eqn.~\ref{eqn:3pulse} and \ref{eqn:gamma_eff} to the data does not uniquely determine $R$ and $\Gamma_{\mathrm{SD}}$ due to their covariance, however, the fit routine does reliably determine their product. In Nd these fits return 
$R\Gamma_{\mathrm{SD}} = \SI{3.5 +- 0.4e6}{\hertz\squared}$, which allows us to extract a limit on the coherence time using Eqn.~\ref{eqn:T2SD} of $T_{2} = \SI{0.60+-0.03}{\milli\second}$. 
Using equations \ref{eqn:R} and \ref{eqn:gammaSD} with the stimulated echo fit, an estimated spin bath temperature of \SI{61}{\milli\kelvin} is obtained (significantly higher than the measured base temperature). Further evidence of the poor thermalisation is seen in the deviation from expected temperature dependence of the echo magnitude below about \SI{100}{\milli\kelvin} (see SI).

Following a similar analysis for the \yb{} sample, the fits return 
$R\Gamma_{SD} = \SI{1.3 +- 0.1e5}{\hertz\squared}$. This results in a spectral diffusion-limited $T_{2}$ of \SI{3.1 +- 0.7}{\milli\second}, consistent with the 2-pulse echo value of \SI{3.38+-0.09}{\milli\second}, and an estimated spin bath temperature of \SI{38}{\milli\kelvin}. We explore the source of spectral diffusion in the \yb{} sample and compute the effect of \y{89} nuclear flip-flops. Following the methods presented in Ref.~\cite{Bottger2006}, described further in the SI, we calculate a spectral diffusion rate and linewidth from \y{89} flip-flops; $R_{Y}\Gamma_{SD,Y} =$\SI{1.05e5}{\hertz\squared} with a limiting $T_{2}=$ \SI{3.48}{\milli\second}. This is in agreement with our measured, $T_{2}$ and we therefore conclude that the low-temperature coherence time is limited by \y{89} nuclear spin flip-flops.

\section{Mitigating Spin decoherence}
In Section III we found the coherence of ~\yb{171} electron spins in ~\yb{nat}-doped YSO at 14~mK to be primarily limited by spectral diffusion from \y{89} nuclear spin flip flops, when measured at \SI{5.04}{\giga\hertz} and 370~mT. We next explore two strategies to further suppress spin decoherence: first, exploiting the substantial anisotropy of the g-tensor and rotating the magnetic field to orientations which give the lowest effective g-factor, and second, exploring the `clock transition' at zero magnetic field for which the first-order sensitivity of the ESR transition frequency goes to zero with respect to magnetic field.

\subsection{Increasing Magnetic Field}
For a given ESR transition frequency, the magnetic field orientation at which the effective g-factor is lowest is that which provides the ESR resonance at the maximum magnetic field. We investigate this approach for $^{171}$Yb using a lower resonator frequency (\SI{2.43}{\giga\hertz}) than that used above so that the resonant fields remain within the range of our vector magnet. 
As the magnetic field is rotated in the D1-D2 plane, the resonant field reaches a peak of \SI{1.2}{\tesla}, however, due to experimental limitations the highest field we studied was \SI{1.07}{\tesla}, at an angle of -131$^\circ$ from D1. Fig.~\ref{fig:fig3-angular_dep}a) shows how $T_{2}$ varies as the field is rotated in the D1-D2 plane. Due to a small misalignment towards the b axis (estimated to be < \SI{0.8}{^{\circ}}), the degeneracy of the two crystal subsites is lifted. 

Considering first the coherence times measured at an angle of around $-88^\circ$ from D1 (approximately along D2), we see that they fall substantially below that predicted due to \y{89} nuclear spin flip-flops. Here, the magnetic field was \SI{154}{}\SI{-174}{\milli\tesla}, less than half that used in Section III due to the lower resonator frequency. In this regime therefore, some electron spin sub-ensembles appear to remain unpolarised at \SI{14}{\milli\kelvin} and contribute to spectral diffusion. As the magnetic field orientation rotates further away from D1, the effective g-factor decreases, demanding a larger magnetic field magnitude to satisfy the resonance condition for $^{171}$Yb --- this increased magnetic field in turn serves to further polarise electron spin sub-ensembles in the bulk, as was achieved in Section III by lowering temperature.
Therefore, approaching the optimum field orientation of $-133^\circ$ offers the twin benefits of i) reducing magnetic field noise by polarising the bath at high field (minimising $R\Gamma_{\rm SD}$), while also reducing the sensitivity, $\nabla_B$\textit{f}, of the central spin to that noise.
Indeed, at $-131^{\circ}$, the coherence time of subsite $a$ was measured to be \SI{6 +- 2}{\milli\second}. 
Both $\nabla_B\textit{f}$ and $R\Gamma_{\rm SD}$ are plotted in Fig.~\ref{fig:fig3-angular_dep}(b)), based on the relationships in Eqs.~\ref{eqn:R} and \ref{eqn:gammaSD} and the spin Hamiltonian (see SI for further details).
%
The resulting model for $T_{2}$ follows the observed data well, including the different times seen for the two subsites. 
We can therefore predict that if the \SI{5.04}{\giga\hertz} resonator were used in this experiment, we would expect a $T_{2}$ of $\SI{25.5}{\milli\second}$ (similar to that measured in \cite{doi:10.1126/sciadv.abj9786}) at $\SI{-132.7}{^{\circ}}$, however this would occur at \SI{3.07}{\tesla}, beyond the capabilities of our experimental setup.

\subsection{Around Zero Magnetic Field}

We have shown above that using large magnetic fields (around \SI{1}{\tesla}) applied at specific orientations can extend the \yb{171} electron spin $T_2$  to several milliseconds.  However, such fields introduce practical challenges when embedding \yb{171} within more complex superconducting circuits, and indeed to integrate with superconducting qubits it is desirable to operate close to zero magnetic field. One of the attractive features of \yb{171} as a quantum memory is the `clock transition' it exhibits at zero field --- here the sensitivity to magnetic field noise goes to zero, to first order, which should reduce impact of spectral diffusion on the spin coherence time.

\begin{figure}
	\includegraphics[width = 0.99\columnwidth]{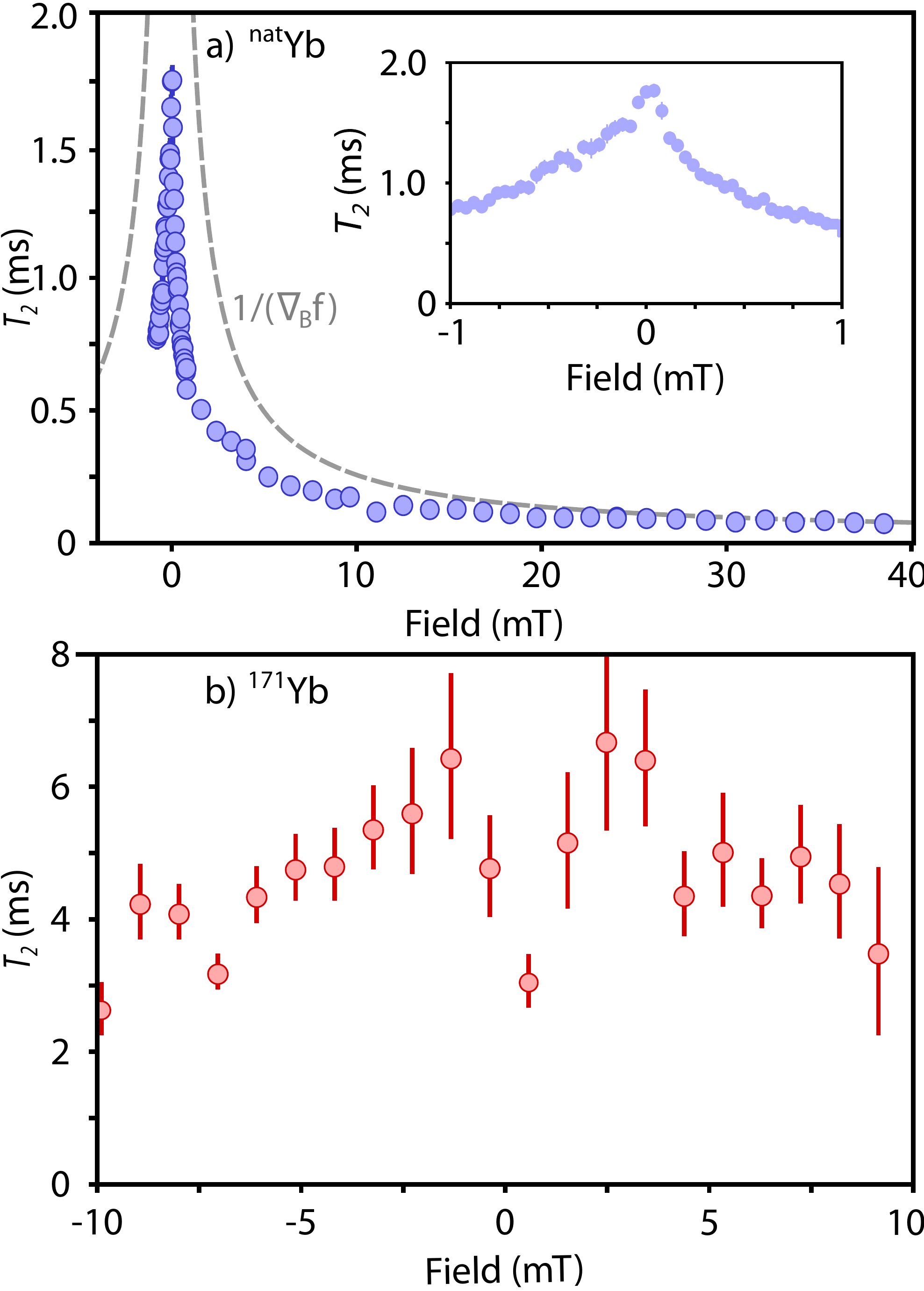}%
	\caption{Field dependence of the coherence time of \yb{171} at low field for both the \yb{nat} doped (a) and isotopically pure \yb{171} doped (b) systems. While the $T_{2}$ increases substantially (to \SI{1.77+-0.06}{\milli\second}) as the the field approaches zero in the \yb{nat} system, the coherence time is not as long as that measured at high field (\SI{6+-2}{\milli\second}). The coherence time deviates away from the trend in decreasing $\nabla_{B}f$ as shown by the grey dashed line. This is due to the presence of an unpolarized spin bath of other \yb{} isotopes at zero field, this is highlighted by the increased coherence in the isotopically pure \yb{171} system (b). Here both a longer maximum $T_{2}$ is measured (\SI{6+-2}{\milli\second}) and a wider region of long coherence is observed.} 
	\label{fig:fig_4}
\end{figure}
%
%
The use of a microresonator to perform ESR at a clock transition at zero magnetic field requires precise control of its resonance frequency to match the \yb{171} zero-field splitting of \SI{2.370}{\giga\hertz}. This challenge is made more acute by the narrowing of the ESR linewidth which occurs around the clock transition. We explore ESR in the low field regime of \yb{nat} doped YSO using a resonator with frequency within 2~MHz of the clock transition (comparable to the spin-resonator coupling strength).

To investigate the impact of the clock transition on spin coherence, we perform magnetic field sweeps along a fixed direction of 49$^{\circ}$ in the D1-D2 plane --- at this field orientation $\nabla_{B}f$ follows a local minimum, leading to the  longest expected coherence times (see SI). Fig.~\ref{fig:fig_4}a) shows how $T_{2}$ increases sharply as the field passes through zero, reaching a maximum of \SI{1.77 +- 0.06}{\milli\second} at \SI{0}{\milli\tesla}. While a strong increase in coherence time (over an order of magnitude) is observed around zero field compared to fields up to 40~mT at this orientation, the increase remains well below that predicted by the decrease in sensitivity to magnetic field noise, $\nabla_Bf$. Indeed, the times measured here at the nominal clock transition at zero field remain several times shorter than those measured at high fields ($\sim1$~T) described above where the effective g-factor drops to 0.1.
We attribute this behaviour to the impact of environmental electron spins that are not only unpolarised, but also become mutually resonant at zero magnetic field, leading to a large increase in the effective magnetic field noise. 
In this \yb{nat}\ sample, isotopes with zero nuclear spin (\yb{I=0}) comprise approximately 70\% of the natural abundance of Yb impurities, providing a large bath of unpolarized electron spins. Approaching zero magnetic field, we conclude that the rate of spectral diffusion increases faster than $\nabla_Bf$ of \yb{171} decreases, limiting the enhancement of $T_2$ (see SI for more details).

\subsection{Isotopic Purification of Yb}

When seeking to exploit \yb{171}\ defect spins in YSO as candidate quantum memory medium, any other isotopes of Yb present in the bulk serve only to introduce additional sources of decoherence, particularly, at the low magnetic fields used to access the clock transition. We next explore the use of isotopically pure \yb{171} doped \yso{}, using an identical resonator design to that used in the \yb{nat} sample with resonant frequency \SI{2.368}{\giga\hertz}. Fig.~\ref{fig:fig_4}b) plots measured $T_{2}$ as a function of magnetic field along the same 49${^\circ}$ orientation as above (the lower resonator frequency here limited the addressable field range). In the absence of other \yb{} isotopes the coherence times are now over an order of magnitude longer in this field range of 0--10~mT, reaching a gentle maximum of \SI{6 +- 1}{ms} at \SI{2.5}{mT} --- the longest \yb{171} $T_{2}$ measured in all the systems we studied here. As the field magnitude reaches $<$ \SI{1}{\milli\tesla}, the measured $T_{2}$ shortens, which we attribute to the impact of \y{89} nuclear spins at low magnetic field. The coherence decay curves themselves exhibit electron spin echo envelope modulation (ESEEM), which occurs when the electron spin interacts with one or more neighbouring nuclear spins and the coherent quantum state is transferred between the two. As the frequency of these oscillations is proportional to the Larmor frequency, the revival of electron spin echo goes to infinity as the field goes to zero. This means that a coherence curve cannot be fitted to the echo decay and so this appears as a decrease in the $T_{2}$, this is a problem present in all systems with a strong hyperfine interaction \cite{simenas_spin_2021}. Another possibility for the dip in coherence at \SI{0}{\milli\tesla} could also be due to the changing behaviour of the nuclear spin bath around the clock transition \cite{PhysRevB.92.161403}, here the \y{89} nuclear spins are no longer in the frozen core and so there is an increase in spectral diffusion. This interaction between the electron spin and the \y{89} nuclear spins is a general issue for operating quantum memory at zero field which effects all rare-earths doped in \yso, motivating studies in other host materials with lower nuclear spin concentrations~\cite{https://doi.org/10.48550/arxiv.2110.04876,doi:10.1073/pnas.2121808119}.

\section{Conclusion}

We used on-chip superconducting resonators to perform pulsed ESR measurements on \nd{145} and \yb{171} in YSO in the high-cooperativity regime $C = $ \numrange{4}{245}. 
Cooling the spin ensemble down to a nominal base temperature of \SI{14}{mK} was shown to extend coherence times by polarising the bath of electron spins in the environment, leading to a $T_2 = \SI{0.41+-0.01}{\milli\second}$ for \nd{145} (likely limited by poor sample thermalisation)
and $T_2 =\SI{3.38+-0.09}{\milli\second}$ for \yb{171}, believed to be limited by spectral diffusion from \y{89} nuclear spins in the host. We explored two routes to suppressing spectral diffusion in \yb{171}.  
The first involved rotating the applied magnetic field to access transitions with low effective g-factor to enable the coherence time of \SI{6+-2}{\milli\second}. The high magnetic field $(>\SI{1}{\tesla})$ serves to polarise the spin bath, reducing the spectral diffusion rate, while the low effective g-factor reduces the sensitivity to spectral diffusion and so the coherence time is further increased. Second, we studied coherence times around the clock transition, showing that the strong reduction in sensitivity of magnetic field noise around zero field is somewhat compensated by a large increase in magnetic field noise from spin flip-flops from other Yb isotopes. Using an isotopically pure \yb{171}:YSO sample we were able to increase the $T_{2}$ to a maximum of \SI{6 +- 1}{\milli\second}.

Overall, based on the understanding we have obtained on decoherence mechanisms in these system, we conclude that \yb{171} offers clear advantages over \nd{145} for spin-based memories in YSO, arising from its lower nuclear spin ($I=1/2$) and thus reduced concentration of non-resonant electron spin in the environment. The longest coherence times for \yb{171} are predicted to be achieved at higher magnetic fields (e.g.\ 25~ms at fields of about 3~T), though these show limited compatibility with superconducting circuits in general. ESEEM from \y{89} is likely to pose a limiting factor in the use of \yb{171} spins at zero magnetic field, however this could be avoided by moving to other crystalline hosts, such as CaWO$_4$ \cite{doi:10.1126/sciadv.abj9786}. Quantum memories schemes typically require high-fidelity control of the spins, which is challenging for bulk-doped rare-earth ion spins such as those studied here --- spatially confining such spins is therefore likely to be advantageous, for example through implantation or lithography.

\begin{acknowledgments}
This work has received funding from the UK Engineering and Physical Sciences Research Council (EPSRC) through the Centre for Doctoral Training in Delivering Quantum Technologies (Grant No.~EP/L015242/1), QUES2T (EP/N015118/1) and the Skills Hub in Quantum Systems Engineering: Innovation in Quantum Business, Applications,
Technology and Engineering (InQuBATE), Grant No.
EP/P510270/1; as well as from the European Research Council (ERC) via LOQOMOTIONS (H2020- EU.1.1., Grant No.\ 771493). We would like to acknowledge support from the UK department for Business Energy and Industrial Strategy (BEIS) through the UK national quantum technologies programme and the ANR MIRESPIN project, Grant No. ANR-19-CE47-0011 of the French Agence Nationale de la Recherche and DGA.

\end{acknowledgments}

\bibliography{bib}

\begin{thebibliography}{37}%
\makeatletter
\providecommand \@ifxundefined [1]{%
 \@ifx{#1\undefined}
}%
\providecommand \@ifnum [1]{%
 \ifnum #1\expandafter \@firstoftwo
 \else \expandafter \@secondoftwo
 \fi
}%
\providecommand \@ifx [1]{%
 \ifx #1\expandafter \@firstoftwo
 \else \expandafter \@secondoftwo
 \fi
}%
\providecommand \natexlab [1]{#1}%
\providecommand \enquote  [1]{``#1''}%
\providecommand \bibnamefont  [1]{#1}%
\providecommand \bibfnamefont [1]{#1}%
\providecommand \citenamefont [1]{#1}%
\providecommand \href@noop [0]{\@secondoftwo}%
\providecommand \href [0]{\begingroup \@sanitize@url \@href}%
\providecommand \@href[1]{\@@startlink{#1}\@@href}%
\providecommand \@@href[1]{\endgroup#1\@@endlink}%
\providecommand \@sanitize@url [0]{\catcode `\\12\catcode `\$12\catcode
  `\&12\catcode `\#12\catcode `\^12\catcode `\_12\catcode `\%12\relax}%
\providecommand \@@startlink[1]{}%
\providecommand \@@endlink[0]{}%
\providecommand \url  [0]{\begingroup\@sanitize@url \@url }%
\providecommand \@url [1]{\endgroup\@href {#1}{\urlprefix }}%
\providecommand \urlprefix  [0]{URL }%
\providecommand \Eprint [0]{\href }%
\providecommand \doibase [0]{https://doi.org/}%
\providecommand \selectlanguage [0]{\@gobble}%
\providecommand \bibinfo  [0]{\@secondoftwo}%
\providecommand \bibfield  [0]{\@secondoftwo}%
\providecommand \translation [1]{[#1]}%
\providecommand \BibitemOpen [0]{}%
\providecommand \bibitemStop [0]{}%
\providecommand \bibitemNoStop [0]{.\EOS\space}%
\providecommand \EOS [0]{\spacefactor3000\relax}%
\providecommand \BibitemShut  [1]{\csname bibitem#1\endcsname}%
\let\auto@bib@innerbib\@empty
\bibitem [{\citenamefont {Thiel}\ \emph {et~al.}(2011)\citenamefont {Thiel},
  \citenamefont {Böttger},\ and\ \citenamefont {Cone}}]{THIEL2011353}%
  \BibitemOpen
  \bibfield  {author} {\bibinfo {author} {\bibfnamefont {C.}~\bibnamefont
  {Thiel}}, \bibinfo {author} {\bibfnamefont {T.}~\bibnamefont {Böttger}},\
  and\ \bibinfo {author} {\bibfnamefont {R.}~\bibnamefont {Cone}},\ }\bibfield
  {title} {\bibinfo {title} {Rare-earth-doped materials for applications in
  quantum information storage and signal processing},\ }\href
  {https://doi.org/https://doi.org/10.1016/j.jlumin.2010.12.015} {\bibfield
  {journal} {\bibinfo  {journal} {Journal of Luminescence}\ }\textbf {\bibinfo
  {volume} {131}},\ \bibinfo {pages} {353 } (\bibinfo {year}
  {2011})}\BibitemShut {NoStop}%
\bibitem [{\citenamefont {Ortu}\ \emph {et~al.}(2018)\citenamefont {Ortu},
  \citenamefont {Tiranov}, \citenamefont {Welinski}, \citenamefont
  {Fr{\"{o}}wis}, \citenamefont {Gisin}, \citenamefont {Ferrier}, \citenamefont
  {Goldner},\ and\ \citenamefont {Afzelius}}]{Ortu2018}%
  \BibitemOpen
  \bibfield  {author} {\bibinfo {author} {\bibfnamefont {A.}~\bibnamefont
  {Ortu}}, \bibinfo {author} {\bibfnamefont {A.}~\bibnamefont {Tiranov}},
  \bibinfo {author} {\bibfnamefont {S.}~\bibnamefont {Welinski}}, \bibinfo
  {author} {\bibfnamefont {F.}~\bibnamefont {Fr{\"{o}}wis}}, \bibinfo {author}
  {\bibfnamefont {N.}~\bibnamefont {Gisin}}, \bibinfo {author} {\bibfnamefont
  {A.}~\bibnamefont {Ferrier}}, \bibinfo {author} {\bibfnamefont
  {P.}~\bibnamefont {Goldner}},\ and\ \bibinfo {author} {\bibfnamefont
  {M.}~\bibnamefont {Afzelius}},\ }\bibfield  {title} {\bibinfo {title}
  {{Simultaneous coherence enhancement of optical and microwave transitions in
  solid-state electronic spins}},\ }\href
  {https://doi.org/10.1038/s41563-018-0138-x} {\bibfield  {journal} {\bibinfo
  {journal} {Nature Materials}\ }\textbf {\bibinfo {volume} {17}},\ \bibinfo
  {pages} {671} (\bibinfo {year} {2018})},\ \Eprint
  {https://arxiv.org/abs/1712.08615} {1712.08615} \BibitemShut {NoStop}%
\bibitem [{\citenamefont {Wolfowicz}\ \emph {et~al.}(2015)\citenamefont
  {Wolfowicz}, \citenamefont {Maier-Flaig}, \citenamefont {Marino},
  \citenamefont {Ferrier}, \citenamefont {Vezin}, \citenamefont {Morton},\ and\
  \citenamefont {Goldner}}]{Wolfowicz2015}%
  \BibitemOpen
  \bibfield  {author} {\bibinfo {author} {\bibfnamefont {G.}~\bibnamefont
  {Wolfowicz}}, \bibinfo {author} {\bibfnamefont {H.}~\bibnamefont
  {Maier-Flaig}}, \bibinfo {author} {\bibfnamefont {R.}~\bibnamefont {Marino}},
  \bibinfo {author} {\bibfnamefont {A.}~\bibnamefont {Ferrier}}, \bibinfo
  {author} {\bibfnamefont {H.}~\bibnamefont {Vezin}}, \bibinfo {author}
  {\bibfnamefont {J.~J.~L.}\ \bibnamefont {Morton}},\ and\ \bibinfo {author}
  {\bibfnamefont {P.}~\bibnamefont {Goldner}},\ }\bibfield  {title} {\bibinfo
  {title} {{Coherent Storage of Microwave Excitations in Rare-Earth Nuclear
  Spins}},\ }\bibfield  {journal} {\bibinfo  {journal} {Physical Review
  Letters}\ }\textbf {\bibinfo {volume} {114}},\ \href
  {https://doi.org/10.1103/PhysRevLett.114.170503}
  {10.1103/PhysRevLett.114.170503} (\bibinfo {year} {2015}),\ \Eprint
  {https://arxiv.org/abs/1412.7298} {1412.7298} \BibitemShut {NoStop}%
\bibitem [{\citenamefont {Li}\ \emph {et~al.}(2020)\citenamefont {Li},
  \citenamefont {Liu}, \citenamefont {Zhou}, \citenamefont {Liu}, \citenamefont
  {Tu}, \citenamefont {Yang}, \citenamefont {Li}, \citenamefont {Ma},
  \citenamefont {Hu}, \citenamefont {Liang}, \citenamefont {Li}, \citenamefont
  {Huang}, \citenamefont {Zhu}, \citenamefont {Li},\ and\ \citenamefont
  {Guo}}]{Li2020}%
  \BibitemOpen
  \bibfield  {author} {\bibinfo {author} {\bibfnamefont {P.-Y.}\ \bibnamefont
  {Li}}, \bibinfo {author} {\bibfnamefont {C.}~\bibnamefont {Liu}}, \bibinfo
  {author} {\bibfnamefont {Z.-Q.}\ \bibnamefont {Zhou}}, \bibinfo {author}
  {\bibfnamefont {X.}~\bibnamefont {Liu}}, \bibinfo {author} {\bibfnamefont
  {T.}~\bibnamefont {Tu}}, \bibinfo {author} {\bibfnamefont {T.-S.}\
  \bibnamefont {Yang}}, \bibinfo {author} {\bibfnamefont {Z.-F.}\ \bibnamefont
  {Li}}, \bibinfo {author} {\bibfnamefont {Y.}~\bibnamefont {Ma}}, \bibinfo
  {author} {\bibfnamefont {J.}~\bibnamefont {Hu}}, \bibinfo {author}
  {\bibfnamefont {P.-J.}\ \bibnamefont {Liang}}, \bibinfo {author}
  {\bibfnamefont {X.}~\bibnamefont {Li}}, \bibinfo {author} {\bibfnamefont
  {J.-Y.}\ \bibnamefont {Huang}}, \bibinfo {author} {\bibfnamefont {T.-X.}\
  \bibnamefont {Zhu}}, \bibinfo {author} {\bibfnamefont {C.-F.}\ \bibnamefont
  {Li}},\ and\ \bibinfo {author} {\bibfnamefont {G.-C.}\ \bibnamefont {Guo}},\
  }\bibfield  {title} {\bibinfo {title} {{Hyperfine Structure and Coherent
  Dynamics of Rare-Earth Spins Explored with Electron-Nuclear Double Resonance
  at Subkelvin Temperatures}},\ }\href
  {https://doi.org/10.1103/PhysRevApplied.13.024080} {\bibfield  {journal}
  {\bibinfo  {journal} {Physical Review Applied}\ }\textbf {\bibinfo {volume}
  {13}},\ \bibinfo {pages} {024080} (\bibinfo {year} {2020})}\BibitemShut
  {NoStop}%
\bibitem [{\citenamefont {Dantec}\ \emph {et~al.}(2021)\citenamefont {Dantec},
  \citenamefont {Rančić}, \citenamefont {Lin}, \citenamefont {Billaud},
  \citenamefont {Ranjan}, \citenamefont {Flanigan}, \citenamefont {Bertaina},
  \citenamefont {Chanelière}, \citenamefont {Goldner}, \citenamefont {Erb},
  \citenamefont {Liu}, \citenamefont {Estève}, \citenamefont {Vion},
  \citenamefont {Flurin},\ and\ \citenamefont
  {Bertet}}]{doi:10.1126/sciadv.abj9786}%
  \BibitemOpen
  \bibfield  {author} {\bibinfo {author} {\bibfnamefont {M.~L.}\ \bibnamefont
  {Dantec}}, \bibinfo {author} {\bibfnamefont {M.}~\bibnamefont {Rančić}},
  \bibinfo {author} {\bibfnamefont {S.}~\bibnamefont {Lin}}, \bibinfo {author}
  {\bibfnamefont {E.}~\bibnamefont {Billaud}}, \bibinfo {author} {\bibfnamefont
  {V.}~\bibnamefont {Ranjan}}, \bibinfo {author} {\bibfnamefont
  {D.}~\bibnamefont {Flanigan}}, \bibinfo {author} {\bibfnamefont
  {S.}~\bibnamefont {Bertaina}}, \bibinfo {author} {\bibfnamefont
  {T.}~\bibnamefont {Chanelière}}, \bibinfo {author} {\bibfnamefont
  {P.}~\bibnamefont {Goldner}}, \bibinfo {author} {\bibfnamefont
  {A.}~\bibnamefont {Erb}}, \bibinfo {author} {\bibfnamefont {R.~B.}\
  \bibnamefont {Liu}}, \bibinfo {author} {\bibfnamefont {D.}~\bibnamefont
  {Estève}}, \bibinfo {author} {\bibfnamefont {D.}~\bibnamefont {Vion}},
  \bibinfo {author} {\bibfnamefont {E.}~\bibnamefont {Flurin}},\ and\ \bibinfo
  {author} {\bibfnamefont {P.}~\bibnamefont {Bertet}},\ }\bibfield  {title}
  {\bibinfo {title} {Twenty-three\&\#x2013;millisecond electron spin coherence
  of erbium ions in a natural-abundance crystal},\ }\href
  {https://doi.org/10.1126/sciadv.abj9786} {\bibfield  {journal} {\bibinfo
  {journal} {Science Advances}\ }\textbf {\bibinfo {volume} {7}},\ \bibinfo
  {pages} {eabj9786} (\bibinfo {year} {2021})},\ \Eprint
  {https://arxiv.org/abs/https://www.science.org/doi/pdf/10.1126/sciadv.abj9786}
  {https://www.science.org/doi/pdf/10.1126/sciadv.abj9786} \BibitemShut
  {NoStop}%
\bibitem [{\citenamefont {Zhong}\ \emph {et~al.}(2015)\citenamefont {Zhong},
  \citenamefont {Hedges}, \citenamefont {Ahlefeldt}, \citenamefont
  {Bartholomew}, \citenamefont {Beavan}, \citenamefont {Wittig}, \citenamefont
  {Longdell},\ and\ \citenamefont {Sellars}}]{Zhong2015}%
  \BibitemOpen
  \bibfield  {author} {\bibinfo {author} {\bibfnamefont {M.}~\bibnamefont
  {Zhong}}, \bibinfo {author} {\bibfnamefont {M.~P.}\ \bibnamefont {Hedges}},
  \bibinfo {author} {\bibfnamefont {R.~L.}\ \bibnamefont {Ahlefeldt}}, \bibinfo
  {author} {\bibfnamefont {J.~G.}\ \bibnamefont {Bartholomew}}, \bibinfo
  {author} {\bibfnamefont {S.~E.}\ \bibnamefont {Beavan}}, \bibinfo {author}
  {\bibfnamefont {S.~M.}\ \bibnamefont {Wittig}}, \bibinfo {author}
  {\bibfnamefont {J.~J.}\ \bibnamefont {Longdell}},\ and\ \bibinfo {author}
  {\bibfnamefont {M.~J.}\ \bibnamefont {Sellars}},\ }\bibfield  {title}
  {\bibinfo {title} {{Optically addressable nuclear spins in a solid with a
  six-hour coherence time}},\ }\href {https://doi.org/10.1038/nature14025}
  {\bibfield  {journal} {\bibinfo  {journal} {Nature}\ }\textbf {\bibinfo
  {volume} {517}},\ \bibinfo {pages} {177} (\bibinfo {year}
  {2015})}\BibitemShut {NoStop}%
\bibitem [{Ran(2017)}]{Rancic2017}%
  \BibitemOpen
  \bibfield  {title} {\bibinfo {title} {{Coherence time of over a second in a
  telecom-compatible quantum memory storage material}},\ }\href
  {https://doi.org/10.1038/nphys4254} {\bibfield  {journal} {\bibinfo
  {journal} {Nature Physics}\ }\textbf {\bibinfo {volume} {14}},\ \bibinfo
  {pages} {50} (\bibinfo {year} {2017})}\BibitemShut {NoStop}%
\bibitem [{\citenamefont {Sangouard}\ \emph {et~al.}(2011)\citenamefont
  {Sangouard}, \citenamefont {Simon}, \citenamefont {de~Riedmatten},\ and\
  \citenamefont {Gisin}}]{Sangouard2011}%
  \BibitemOpen
  \bibfield  {author} {\bibinfo {author} {\bibfnamefont {N.}~\bibnamefont
  {Sangouard}}, \bibinfo {author} {\bibfnamefont {C.}~\bibnamefont {Simon}},
  \bibinfo {author} {\bibfnamefont {H.}~\bibnamefont {de~Riedmatten}},\ and\
  \bibinfo {author} {\bibfnamefont {N.}~\bibnamefont {Gisin}},\ }\bibfield
  {title} {\bibinfo {title} {{Quantum repeaters based on atomic ensembles and
  linear optics}},\ }\href {https://doi.org/10.1103/RevModPhys.83.33}
  {\bibfield  {journal} {\bibinfo  {journal} {Reviews of Modern Physics}\
  }\textbf {\bibinfo {volume} {83}},\ \bibinfo {pages} {33} (\bibinfo {year}
  {2011})},\ \Eprint {https://arxiv.org/abs/0906.2699} {0906.2699} \BibitemShut
  {NoStop}%
\bibitem [{\citenamefont {Probst}\ \emph {et~al.}(2015)\citenamefont {Probst},
  \citenamefont {Rotzinger}, \citenamefont {Ustinov},\ and\ \citenamefont
  {Bushev}}]{Probst2015}%
  \BibitemOpen
  \bibfield  {author} {\bibinfo {author} {\bibfnamefont {S.}~\bibnamefont
  {Probst}}, \bibinfo {author} {\bibfnamefont {H.}~\bibnamefont {Rotzinger}},
  \bibinfo {author} {\bibfnamefont {A.~V.}\ \bibnamefont {Ustinov}},\ and\
  \bibinfo {author} {\bibfnamefont {P.~A.}\ \bibnamefont {Bushev}},\ }\bibfield
   {title} {\bibinfo {title} {{Microwave multimode memory with an erbium spin
  ensemble}},\ }\href {https://doi.org/10.1103/PhysRevB.92.014421} {\bibfield
  {journal} {\bibinfo  {journal} {Physical Review B}\ }\textbf {\bibinfo
  {volume} {92}},\ \bibinfo {pages} {014421} (\bibinfo {year}
  {2015})}\BibitemShut {NoStop}%
\bibitem [{\citenamefont {Zhong}\ \emph {et~al.}(2017)\citenamefont {Zhong},
  \citenamefont {Kindem}, \citenamefont {Bartholomew}, \citenamefont {Rochman},
  \citenamefont {Craiciu}, \citenamefont {Miyazono}, \citenamefont
  {Bettinelli}, \citenamefont {Cavalli}, \citenamefont {Verma}, \citenamefont
  {Nam}, \citenamefont {Marsili}, \citenamefont {Shaw}, \citenamefont {Beyer},\
  and\ \citenamefont {Faraon}}]{doi:10.1126/science.aan5959}%
  \BibitemOpen
  \bibfield  {author} {\bibinfo {author} {\bibfnamefont {T.}~\bibnamefont
  {Zhong}}, \bibinfo {author} {\bibfnamefont {J.~M.}\ \bibnamefont {Kindem}},
  \bibinfo {author} {\bibfnamefont {J.~G.}\ \bibnamefont {Bartholomew}},
  \bibinfo {author} {\bibfnamefont {J.}~\bibnamefont {Rochman}}, \bibinfo
  {author} {\bibfnamefont {I.}~\bibnamefont {Craiciu}}, \bibinfo {author}
  {\bibfnamefont {E.}~\bibnamefont {Miyazono}}, \bibinfo {author}
  {\bibfnamefont {M.}~\bibnamefont {Bettinelli}}, \bibinfo {author}
  {\bibfnamefont {E.}~\bibnamefont {Cavalli}}, \bibinfo {author} {\bibfnamefont
  {V.}~\bibnamefont {Verma}}, \bibinfo {author} {\bibfnamefont {S.~W.}\
  \bibnamefont {Nam}}, \bibinfo {author} {\bibfnamefont {F.}~\bibnamefont
  {Marsili}}, \bibinfo {author} {\bibfnamefont {M.~D.}\ \bibnamefont {Shaw}},
  \bibinfo {author} {\bibfnamefont {A.~D.}\ \bibnamefont {Beyer}},\ and\
  \bibinfo {author} {\bibfnamefont {A.}~\bibnamefont {Faraon}},\ }\bibfield
  {title} {\bibinfo {title} {Nanophotonic rare-earth quantum memory with
  optically controlled retrieval},\ }\href
  {https://doi.org/10.1126/science.aan5959} {\bibfield  {journal} {\bibinfo
  {journal} {Science}\ }\textbf {\bibinfo {volume} {357}},\ \bibinfo {pages}
  {1392} (\bibinfo {year} {2017})},\ \Eprint
  {https://arxiv.org/abs/https://www.science.org/doi/pdf/10.1126/science.aan5959}
  {https://www.science.org/doi/pdf/10.1126/science.aan5959} \BibitemShut
  {NoStop}%
\bibitem [{\citenamefont {Laplane}\ \emph {et~al.}(2017)\citenamefont
  {Laplane}, \citenamefont {Jobez}, \citenamefont {Etesse}, \citenamefont
  {Gisin},\ and\ \citenamefont {Afzelius}}]{PhysRevLett.118.210501}%
  \BibitemOpen
  \bibfield  {author} {\bibinfo {author} {\bibfnamefont {C.}~\bibnamefont
  {Laplane}}, \bibinfo {author} {\bibfnamefont {P.}~\bibnamefont {Jobez}},
  \bibinfo {author} {\bibfnamefont {J.}~\bibnamefont {Etesse}}, \bibinfo
  {author} {\bibfnamefont {N.}~\bibnamefont {Gisin}},\ and\ \bibinfo {author}
  {\bibfnamefont {M.}~\bibnamefont {Afzelius}},\ }\bibfield  {title} {\bibinfo
  {title} {Multimode and long-lived quantum correlations between photons and
  spins in a crystal},\ }\href {https://doi.org/10.1103/PhysRevLett.118.210501}
  {\bibfield  {journal} {\bibinfo  {journal} {Phys. Rev. Lett.}\ }\textbf
  {\bibinfo {volume} {118}},\ \bibinfo {pages} {210501} (\bibinfo {year}
  {2017})}\BibitemShut {NoStop}%
\bibitem [{\citenamefont {Ruskuc}\ \emph {et~al.}(2022)\citenamefont {Ruskuc},
  \citenamefont {Wu}, \citenamefont {Rochman}, \citenamefont {Choi},\ and\
  \citenamefont {Faraon}}]{ruskuc_nuclear_2022}%
  \BibitemOpen
  \bibfield  {author} {\bibinfo {author} {\bibfnamefont {A.}~\bibnamefont
  {Ruskuc}}, \bibinfo {author} {\bibfnamefont {C.-J.}\ \bibnamefont {Wu}},
  \bibinfo {author} {\bibfnamefont {J.}~\bibnamefont {Rochman}}, \bibinfo
  {author} {\bibfnamefont {J.}~\bibnamefont {Choi}},\ and\ \bibinfo {author}
  {\bibfnamefont {A.}~\bibnamefont {Faraon}},\ }\bibfield  {title} {\bibinfo
  {title} {Nuclear spin-wave quantum register for a solid-state qubit},\ }\href
  {https://doi.org/10.1038/s41586-021-04293-6} {\bibfield  {journal} {\bibinfo
  {journal} {Nature}\ }\textbf {\bibinfo {volume} {602}},\ \bibinfo {pages}
  {408} (\bibinfo {year} {2022})}\BibitemShut {NoStop}%
\bibitem [{\citenamefont {Williamson}\ \emph {et~al.}(2014)\citenamefont
  {Williamson}, \citenamefont {Chen},\ and\ \citenamefont
  {Longdell}}]{Williamson2014}%
  \BibitemOpen
  \bibfield  {author} {\bibinfo {author} {\bibfnamefont {L.~A.}\ \bibnamefont
  {Williamson}}, \bibinfo {author} {\bibfnamefont {Y.~H.}\ \bibnamefont
  {Chen}},\ and\ \bibinfo {author} {\bibfnamefont {J.~J.}\ \bibnamefont
  {Longdell}},\ }\bibfield  {title} {\bibinfo {title} {{Magneto-optic modulator
  with unit quantum efficiency}},\ }\href
  {https://doi.org/10.1103/PhysRevLett.113.203601} {\bibfield  {journal}
  {\bibinfo  {journal} {Physical Review Letters}\ }\textbf {\bibinfo {volume}
  {113}},\ \bibinfo {pages} {1} (\bibinfo {year} {2014})},\ \Eprint
  {https://arxiv.org/abs/1403.1608} {1403.1608} \BibitemShut {NoStop}%
\bibitem [{\citenamefont {Fernandez-Gonzalvo}\ \emph
  {et~al.}(2015)\citenamefont {Fernandez-Gonzalvo}, \citenamefont {Chen},
  \citenamefont {Yin}, \citenamefont {Rogge},\ and\ \citenamefont
  {Longdell}}]{Fernandez-Gonzalvo2015}%
  \BibitemOpen
  \bibfield  {author} {\bibinfo {author} {\bibfnamefont {X.}~\bibnamefont
  {Fernandez-Gonzalvo}}, \bibinfo {author} {\bibfnamefont {Y.~H.}\ \bibnamefont
  {Chen}}, \bibinfo {author} {\bibfnamefont {C.}~\bibnamefont {Yin}}, \bibinfo
  {author} {\bibfnamefont {S.}~\bibnamefont {Rogge}},\ and\ \bibinfo {author}
  {\bibfnamefont {J.~J.}\ \bibnamefont {Longdell}},\ }\bibfield  {title}
  {\bibinfo {title} {{Coherent frequency up-conversion of microwaves to the
  optical telecommunications band in an Er:YSO crystal}},\ }\href
  {https://doi.org/10.1103/PhysRevA.92.062313} {\bibfield  {journal} {\bibinfo
  {journal} {Physical Review A - Atomic, Molecular, and Optical Physics}\
  }\textbf {\bibinfo {volume} {92}},\ \bibinfo {pages} {1} (\bibinfo {year}
  {2015})},\ \Eprint {https://arxiv.org/abs/1501.02014} {1501.02014}
  \BibitemShut {NoStop}%
\bibitem [{\citenamefont {Bartholomew}\ \emph {et~al.}(2020)\citenamefont
  {Bartholomew}, \citenamefont {Rochman}, \citenamefont {Xie}, \citenamefont
  {Kindem}, \citenamefont {Ruskuc}, \citenamefont {Craiciu}, \citenamefont
  {Lei},\ and\ \citenamefont {Faraon}}]{bartholomew_-chip_2020}%
  \BibitemOpen
  \bibfield  {author} {\bibinfo {author} {\bibfnamefont {J.~G.}\ \bibnamefont
  {Bartholomew}}, \bibinfo {author} {\bibfnamefont {J.}~\bibnamefont
  {Rochman}}, \bibinfo {author} {\bibfnamefont {T.}~\bibnamefont {Xie}},
  \bibinfo {author} {\bibfnamefont {J.~M.}\ \bibnamefont {Kindem}}, \bibinfo
  {author} {\bibfnamefont {A.}~\bibnamefont {Ruskuc}}, \bibinfo {author}
  {\bibfnamefont {I.}~\bibnamefont {Craiciu}}, \bibinfo {author} {\bibfnamefont
  {M.}~\bibnamefont {Lei}},\ and\ \bibinfo {author} {\bibfnamefont
  {A.}~\bibnamefont {Faraon}},\ }\bibfield  {title} {\bibinfo {title} {On-chip
  coherent microwave-to-optical transduction mediated by ytterbium in {YVO4}},\
  }\href {https://doi.org/10.1038/s41467-020-16996-x} {\bibfield  {journal}
  {\bibinfo  {journal} {Nature Communications}\ }\textbf {\bibinfo {volume}
  {11}},\ \bibinfo {pages} {3266} (\bibinfo {year} {2020})}\BibitemShut
  {NoStop}%
\bibitem [{\citenamefont {Tavis}\ and\ \citenamefont
  {Cummings}(1968)}]{Tavis1968}%
  \BibitemOpen
  \bibfield  {author} {\bibinfo {author} {\bibfnamefont {M.}~\bibnamefont
  {Tavis}}\ and\ \bibinfo {author} {\bibfnamefont {F.~W.}\ \bibnamefont
  {Cummings}},\ }\bibfield  {title} {\bibinfo {title} {{Exact solution for an
  N-molecule-radiation-field Hamiltonian}},\ }\bibfield  {journal} {\bibinfo
  {journal} {Physical Review}\ }\textbf {\bibinfo {volume} {170}},\ \href
  {https://doi.org/10.1103/PhysRev.170.379} {10.1103/PhysRev.170.379} (\bibinfo
  {year} {1968})\BibitemShut {NoStop}%
\bibitem [{\citenamefont {Afzelius}\ \emph {et~al.}(2013)\citenamefont
  {Afzelius}, \citenamefont {Sangouard}, \citenamefont {Johansson},
  \citenamefont {Staudt},\ and\ \citenamefont {Wilson}}]{afzelius2013proposal}%
  \BibitemOpen
  \bibfield  {author} {\bibinfo {author} {\bibfnamefont {M.}~\bibnamefont
  {Afzelius}}, \bibinfo {author} {\bibfnamefont {N.}~\bibnamefont {Sangouard}},
  \bibinfo {author} {\bibfnamefont {G.}~\bibnamefont {Johansson}}, \bibinfo
  {author} {\bibfnamefont {M.}~\bibnamefont {Staudt}},\ and\ \bibinfo {author}
  {\bibfnamefont {C.}~\bibnamefont {Wilson}},\ }\bibfield  {title} {\bibinfo
  {title} {Proposal for a coherent quantum memory for propagating microwave
  photons},\ }\href@noop {} {\bibfield  {journal} {\bibinfo  {journal} {New
  Journal of Physics}\ }\textbf {\bibinfo {volume} {15}},\ \bibinfo {pages}
  {065008} (\bibinfo {year} {2013})}\BibitemShut {NoStop}%
\bibitem [{\citenamefont {Julsgaard}\ \emph {et~al.}(2013)\citenamefont
  {Julsgaard}, \citenamefont {Grezes}, \citenamefont {Bertet},\ and\
  \citenamefont {M{\o}lmer}}]{julsgaard2013quantum}%
  \BibitemOpen
  \bibfield  {author} {\bibinfo {author} {\bibfnamefont {B.}~\bibnamefont
  {Julsgaard}}, \bibinfo {author} {\bibfnamefont {C.}~\bibnamefont {Grezes}},
  \bibinfo {author} {\bibfnamefont {P.}~\bibnamefont {Bertet}},\ and\ \bibinfo
  {author} {\bibfnamefont {K.}~\bibnamefont {M{\o}lmer}},\ }\bibfield  {title}
  {\bibinfo {title} {Quantum memory for microwave photons in an inhomogeneously
  broadened spin ensemble},\ }\href@noop {} {\bibfield  {journal} {\bibinfo
  {journal} {Physical Review Letters}\ }\textbf {\bibinfo {volume} {110}},\
  \bibinfo {pages} {250503} (\bibinfo {year} {2013})}\BibitemShut {NoStop}%
\bibitem [{\citenamefont {Grezes}\ \emph {et~al.}(2016)\citenamefont {Grezes},
  \citenamefont {Kubo}, \citenamefont {Julsgaard}, \citenamefont {Umeda},
  \citenamefont {Isoya}, \citenamefont {Sumiya}, \citenamefont {Abe},
  \citenamefont {Onoda}, \citenamefont {Ohshima}, \citenamefont {Nakamura},
  \citenamefont {Diniz}, \citenamefont {Auffeves}, \citenamefont {Jacques},
  \citenamefont {Roch}, \citenamefont {Vion}, \citenamefont {Esteve},
  \citenamefont {Moelmer},\ and\ \citenamefont {Bertet}}]{grezes2016towards}%
  \BibitemOpen
  \bibfield  {author} {\bibinfo {author} {\bibfnamefont {C.}~\bibnamefont
  {Grezes}}, \bibinfo {author} {\bibfnamefont {Y.}~\bibnamefont {Kubo}},
  \bibinfo {author} {\bibfnamefont {B.}~\bibnamefont {Julsgaard}}, \bibinfo
  {author} {\bibfnamefont {T.}~\bibnamefont {Umeda}}, \bibinfo {author}
  {\bibfnamefont {J.}~\bibnamefont {Isoya}}, \bibinfo {author} {\bibfnamefont
  {H.}~\bibnamefont {Sumiya}}, \bibinfo {author} {\bibfnamefont
  {H.}~\bibnamefont {Abe}}, \bibinfo {author} {\bibfnamefont {S.}~\bibnamefont
  {Onoda}}, \bibinfo {author} {\bibfnamefont {T.}~\bibnamefont {Ohshima}},
  \bibinfo {author} {\bibfnamefont {K.}~\bibnamefont {Nakamura}}, \bibinfo
  {author} {\bibfnamefont {I.}~\bibnamefont {Diniz}}, \bibinfo {author}
  {\bibfnamefont {A.}~\bibnamefont {Auffeves}}, \bibinfo {author}
  {\bibfnamefont {V.}~\bibnamefont {Jacques}}, \bibinfo {author} {\bibfnamefont
  {J.-F.}\ \bibnamefont {Roch}}, \bibinfo {author} {\bibfnamefont
  {D.}~\bibnamefont {Vion}}, \bibinfo {author} {\bibfnamefont {D.}~\bibnamefont
  {Esteve}}, \bibinfo {author} {\bibfnamefont {K.}~\bibnamefont {Moelmer}},\
  and\ \bibinfo {author} {\bibfnamefont {P.}~\bibnamefont {Bertet}},\
  }\bibfield  {title} {\bibinfo {title} {Towards a spin-ensemble quantum memory
  for superconducting qubits},\ }\href
  {https://doi.org/https://doi.org/10.1016/j.crhy.2016.07.006} {\bibfield
  {journal} {\bibinfo  {journal} {Comptes Rendus Physique}\ }\textbf {\bibinfo
  {volume} {17}},\ \bibinfo {pages} {693 } (\bibinfo {year}
  {2016})}\BibitemShut {NoStop}%
\bibitem [{\citenamefont {B{\"{o}}ttger}\ \emph {et~al.}(2006)\citenamefont
  {B{\"{o}}ttger}, \citenamefont {Thiel}, \citenamefont {Sun},\ and\
  \citenamefont {Cone}}]{Bottger2006}%
  \BibitemOpen
  \bibfield  {author} {\bibinfo {author} {\bibfnamefont {T.}~\bibnamefont
  {B{\"{o}}ttger}}, \bibinfo {author} {\bibfnamefont {C.~W.}\ \bibnamefont
  {Thiel}}, \bibinfo {author} {\bibfnamefont {Y.}~\bibnamefont {Sun}},\ and\
  \bibinfo {author} {\bibfnamefont {R.~L.}\ \bibnamefont {Cone}},\ }\bibfield
  {title} {\bibinfo {title} {Optical decoherence and spectral diffusion at 1.5
  $\mu$m in er$^{3+}$: Y$_{2}$ sio$_{5}$ versus magnetic field, temperature,
  and er$^{3+}$ concentration},\ }\href
  {https://doi.org/10.1103/PhysRevB.73.075101} {\bibfield  {journal} {\bibinfo
  {journal} {Physical Review B - Condensed Matter and Materials Physics}\
  }\textbf {\bibinfo {volume} {73}},\ \bibinfo {pages} {1} (\bibinfo {year}
  {2006})}\BibitemShut {NoStop}%
\bibitem [{\citenamefont {Salikhov}\ \emph {et~al.}(1981)\citenamefont
  {Salikhov}, \citenamefont {Dzuba},\ and\ \citenamefont
  {Raitsimring}}]{SALIKHOV1981255}%
  \BibitemOpen
  \bibfield  {author} {\bibinfo {author} {\bibfnamefont {K.}~\bibnamefont
  {Salikhov}}, \bibinfo {author} {\bibfnamefont {S.}~\bibnamefont {Dzuba}},\
  and\ \bibinfo {author} {\bibfnamefont {A.}~\bibnamefont {Raitsimring}},\
  }\bibfield  {title} {\bibinfo {title} {The theory of electron spin-echo
  signal decay resulting from dipole-dipole interactions between paramagnetic
  centers in solids},\ }\href
  {https://doi.org/https://doi.org/10.1016/0022-2364(81)90216-X} {\bibfield
  {journal} {\bibinfo  {journal} {Journal of Magnetic Resonance (1969)}\
  }\textbf {\bibinfo {volume} {42}},\ \bibinfo {pages} {255} (\bibinfo {year}
  {1981})}\BibitemShut {NoStop}%
\bibitem [{\citenamefont {Dold}\ \emph {et~al.}(2019)\citenamefont {Dold},
  \citenamefont {Zollitsch}, \citenamefont {O'Sullivan}, \citenamefont
  {Welinski}, \citenamefont {Ferrier}, \citenamefont {Goldner}, \citenamefont
  {de~Graaf}, \citenamefont {Lindstr{\"{o}}m},\ and\ \citenamefont
  {Morton}}]{Dold2019}%
  \BibitemOpen
  \bibfield  {author} {\bibinfo {author} {\bibfnamefont {G.}~\bibnamefont
  {Dold}}, \bibinfo {author} {\bibfnamefont {C.~W.}\ \bibnamefont {Zollitsch}},
  \bibinfo {author} {\bibfnamefont {J.}~\bibnamefont {O'Sullivan}}, \bibinfo
  {author} {\bibfnamefont {S.}~\bibnamefont {Welinski}}, \bibinfo {author}
  {\bibfnamefont {A.}~\bibnamefont {Ferrier}}, \bibinfo {author} {\bibfnamefont
  {P.}~\bibnamefont {Goldner}}, \bibinfo {author} {\bibfnamefont
  {S.}~\bibnamefont {de~Graaf}}, \bibinfo {author} {\bibfnamefont
  {T.}~\bibnamefont {Lindstr{\"{o}}m}},\ and\ \bibinfo {author} {\bibfnamefont
  {J.~J.}\ \bibnamefont {Morton}},\ }\bibfield  {title} {\bibinfo {title}
  {{High-Cooperativity Coupling of a Rare-Earth Spin Ensemble to a
  Superconducting Resonator Using Yttrium Orthosilicate as a Substrate}},\
  }\href {https://doi.org/10.1103/PhysRevApplied.11.054082} {\bibfield
  {journal} {\bibinfo  {journal} {Physical Review Applied}\ }\textbf {\bibinfo
  {volume} {11}},\ \bibinfo {pages} {054082} (\bibinfo {year}
  {2019})}\BibitemShut {NoStop}%
\bibitem [{\citenamefont {Fraval}\ \emph {et~al.}(2004)\citenamefont {Fraval},
  \citenamefont {Sellars},\ and\ \citenamefont
  {Longdell}}]{PhysRevLett.92.077601}%
  \BibitemOpen
  \bibfield  {author} {\bibinfo {author} {\bibfnamefont {E.}~\bibnamefont
  {Fraval}}, \bibinfo {author} {\bibfnamefont {M.~J.}\ \bibnamefont
  {Sellars}},\ and\ \bibinfo {author} {\bibfnamefont {J.~J.}\ \bibnamefont
  {Longdell}},\ }\bibfield  {title} {\bibinfo {title} {Method of extending
  hyperfine coherence times in
  ${\mathrm{p}\mathrm{r}}^{3+}\ensuremath{\mathbin:}{\mathrm{y}}_{2}{\mathrm{s}\mathrm{i}\mathrm{o}}_{5}$},\
  }\href {https://doi.org/10.1103/PhysRevLett.92.077601} {\bibfield  {journal}
  {\bibinfo  {journal} {Phys. Rev. Lett.}\ }\textbf {\bibinfo {volume} {92}},\
  \bibinfo {pages} {077601} (\bibinfo {year} {2004})}\BibitemShut {NoStop}%
\bibitem [{\citenamefont {Cruzeiro}\ \emph {et~al.}(2017)\citenamefont
  {Cruzeiro}, \citenamefont {Tiranov}, \citenamefont {Usmani}, \citenamefont
  {Laplane}, \citenamefont {Lavoie}, \citenamefont {Ferrier}, \citenamefont
  {Goldner}, \citenamefont {Gisin},\ and\ \citenamefont
  {Afzelius}}]{Cruzeiro2016}%
  \BibitemOpen
  \bibfield  {author} {\bibinfo {author} {\bibfnamefont {E.~Z.}\ \bibnamefont
  {Cruzeiro}}, \bibinfo {author} {\bibfnamefont {A.}~\bibnamefont {Tiranov}},
  \bibinfo {author} {\bibfnamefont {I.}~\bibnamefont {Usmani}}, \bibinfo
  {author} {\bibfnamefont {C.}~\bibnamefont {Laplane}}, \bibinfo {author}
  {\bibfnamefont {J.}~\bibnamefont {Lavoie}}, \bibinfo {author} {\bibfnamefont
  {A.}~\bibnamefont {Ferrier}}, \bibinfo {author} {\bibfnamefont
  {P.}~\bibnamefont {Goldner}}, \bibinfo {author} {\bibfnamefont
  {N.}~\bibnamefont {Gisin}},\ and\ \bibinfo {author} {\bibfnamefont
  {M.}~\bibnamefont {Afzelius}},\ }\bibfield  {title} {\bibinfo {title}
  {{Spectral hole lifetimes and spin population relaxation dynamics in
  neodymium-doped yttrium orthosilicate}},\ }\href
  {https://doi.org/10.1103/PhysRevB.95.205119} {\bibfield  {journal} {\bibinfo
  {journal} {Physical Review B}\ }\textbf {\bibinfo {volume} {95}},\ \bibinfo
  {pages} {205119} (\bibinfo {year} {2017})},\ \Eprint
  {https://arxiv.org/abs/1611.05444} {1611.05444} \BibitemShut {NoStop}%
\bibitem [{\citenamefont {Schweiger}\ and\ \citenamefont
  {Jeschke}(2001)}]{Schweiger2001}%
  \BibitemOpen
  \bibfield  {author} {\bibinfo {author} {\bibfnamefont {A.}~\bibnamefont
  {Schweiger}}\ and\ \bibinfo {author} {\bibfnamefont {G.}~\bibnamefont
  {Jeschke}},\ }\href {https://doi.org/10.1002/jctb.936} {\emph {\bibinfo
  {title} {Principles of Pulse Electron Paramagnetic Resonance}}}\ (\bibinfo
  {publisher} {Oxford University Press},\ \bibinfo {year} {2001})\BibitemShut
  {NoStop}%
\bibitem [{\citenamefont {Tyryshkin}\ \emph {et~al.}(2011)\citenamefont
  {Tyryshkin}, \citenamefont {Tojo}, \citenamefont {Morton}, \citenamefont
  {Riemann}, \citenamefont {Abrosimov}, \citenamefont {Becker}, \citenamefont
  {Pohl}, \citenamefont {Schenkel}, \citenamefont {Thewalt}, \citenamefont
  {Itoh},\ and\ \citenamefont {Lyon}}]{Tyryshkin2011a}%
  \BibitemOpen
  \bibfield  {author} {\bibinfo {author} {\bibfnamefont {A.~M.}\ \bibnamefont
  {Tyryshkin}}, \bibinfo {author} {\bibfnamefont {S.}~\bibnamefont {Tojo}},
  \bibinfo {author} {\bibfnamefont {J.~J.~L.}\ \bibnamefont {Morton}}, \bibinfo
  {author} {\bibfnamefont {H.}~\bibnamefont {Riemann}}, \bibinfo {author}
  {\bibfnamefont {N.~V.}\ \bibnamefont {Abrosimov}}, \bibinfo {author}
  {\bibfnamefont {P.}~\bibnamefont {Becker}}, \bibinfo {author} {\bibfnamefont
  {H.-J.}\ \bibnamefont {Pohl}}, \bibinfo {author} {\bibfnamefont
  {T.}~\bibnamefont {Schenkel}}, \bibinfo {author} {\bibfnamefont {M.~L.~W.}\
  \bibnamefont {Thewalt}}, \bibinfo {author} {\bibfnamefont {K.~M.}\
  \bibnamefont {Itoh}},\ and\ \bibinfo {author} {\bibfnamefont {S.~A.}\
  \bibnamefont {Lyon}},\ }\bibfield  {title} {\bibinfo {title} {{Electron spin
  coherence exceeding seconds in high-purity silicon}},\ }\href
  {https://doi.org/10.1038/nmat3182} {\bibfield  {journal} {\bibinfo  {journal}
  {Nature Materials}\ }\textbf {\bibinfo {volume} {11}},\ \bibinfo {pages}
  {143} (\bibinfo {year} {2011})}\BibitemShut {NoStop}%
\bibitem [{\citenamefont {Lim}\ \emph {et~al.}(2017)\citenamefont {Lim},
  \citenamefont {Welinski}, \citenamefont {Ferrier}, \citenamefont {Goldner},\
  and\ \citenamefont {Morton}}]{Lim2017}%
  \BibitemOpen
  \bibfield  {author} {\bibinfo {author} {\bibfnamefont {H.-J.}\ \bibnamefont
  {Lim}}, \bibinfo {author} {\bibfnamefont {S.}~\bibnamefont {Welinski}},
  \bibinfo {author} {\bibfnamefont {A.}~\bibnamefont {Ferrier}}, \bibinfo
  {author} {\bibfnamefont {P.}~\bibnamefont {Goldner}},\ and\ \bibinfo {author}
  {\bibfnamefont {J.~J.~L.}\ \bibnamefont {Morton}},\ }\bibfield  {title}
  {\bibinfo {title} {{Coherent spin dynamics of ytterbium ions in yttrium
  orthosilicate}},\ }\href {https://doi.org/10.1103/PhysRevB.97.064409}
  {\bibfield  {journal} {\bibinfo  {journal} {Physical Review B}\ }\textbf
  {\bibinfo {volume} {97}},\ \bibinfo {pages} {064409} (\bibinfo {year}
  {2017})},\ \Eprint {https://arxiv.org/abs/1712.00435} {1712.00435}
  \BibitemShut {NoStop}%
\bibitem [{\citenamefont {Takahashi}\ \emph {et~al.}(2008)\citenamefont
  {Takahashi}, \citenamefont {Hanson}, \citenamefont {{Van Tol}}, \citenamefont
  {Sherwin},\ and\ \citenamefont {Awschalom}}]{Takahashi2008}%
  \BibitemOpen
  \bibfield  {author} {\bibinfo {author} {\bibfnamefont {S.}~\bibnamefont
  {Takahashi}}, \bibinfo {author} {\bibfnamefont {R.}~\bibnamefont {Hanson}},
  \bibinfo {author} {\bibfnamefont {J.}~\bibnamefont {{Van Tol}}}, \bibinfo
  {author} {\bibfnamefont {M.~S.}\ \bibnamefont {Sherwin}},\ and\ \bibinfo
  {author} {\bibfnamefont {D.~D.}\ \bibnamefont {Awschalom}},\ }\bibfield
  {title} {\bibinfo {title} {{Quenching spin decoherence in diamond through
  spin bath polarization}},\ }\bibfield  {journal} {\bibinfo  {journal}
  {Physical Review Letters}\ }\textbf {\bibinfo {volume} {101}},\ \href
  {https://doi.org/10.1053/j.sodo.2011.07.007} {10.1053/j.sodo.2011.07.007}
  (\bibinfo {year} {2008}),\ \Eprint {https://arxiv.org/abs/0804.1537}
  {0804.1537} \BibitemShut {NoStop}%
\bibitem [{\citenamefont {Rančić}\ \emph {et~al.}(2022)\citenamefont
  {Rančić}, \citenamefont {Dantec}, \citenamefont {Lin}, \citenamefont
  {Bertaina}, \citenamefont {Chanelière}, \citenamefont {Serrano},
  \citenamefont {Goldner}, \citenamefont {Liu}, \citenamefont {Flurin},
  \citenamefont {Estève}, \citenamefont {Vion},\ and\ \citenamefont
  {Bertet}}]{rancic_electron-spin_2022}%
  \BibitemOpen
  \bibfield  {author} {\bibinfo {author} {\bibfnamefont {M.}~\bibnamefont
  {Rančić}}, \bibinfo {author} {\bibfnamefont {M.~L.}\ \bibnamefont
  {Dantec}}, \bibinfo {author} {\bibfnamefont {S.}~\bibnamefont {Lin}},
  \bibinfo {author} {\bibfnamefont {S.}~\bibnamefont {Bertaina}}, \bibinfo
  {author} {\bibfnamefont {T.}~\bibnamefont {Chanelière}}, \bibinfo {author}
  {\bibfnamefont {D.}~\bibnamefont {Serrano}}, \bibinfo {author} {\bibfnamefont
  {P.}~\bibnamefont {Goldner}}, \bibinfo {author} {\bibfnamefont {R.~B.}\
  \bibnamefont {Liu}}, \bibinfo {author} {\bibfnamefont {E.}~\bibnamefont
  {Flurin}}, \bibinfo {author} {\bibfnamefont {D.}~\bibnamefont {Estève}},
  \bibinfo {author} {\bibfnamefont {D.}~\bibnamefont {Vion}},\ and\ \bibinfo
  {author} {\bibfnamefont {P.}~\bibnamefont {Bertet}},\ }\href
  {http://arxiv.org/abs/2203.15012} {\bibinfo {title} {Electron-spin spectral
  diffusion in an erbium doped crystal at millikelvin temperatures}} (\bibinfo
  {year} {2022})\BibitemShut {NoStop}%
\bibitem [{\citenamefont {Sun}\ \emph {et~al.}(2008)\citenamefont {Sun},
  \citenamefont {B{\"{o}}ttger}, \citenamefont {Thiel},\ and\ \citenamefont
  {Cone}}]{Sun2008}%
  \BibitemOpen
  \bibfield  {author} {\bibinfo {author} {\bibfnamefont {Y.}~\bibnamefont
  {Sun}}, \bibinfo {author} {\bibfnamefont {T.}~\bibnamefont {B{\"{o}}ttger}},
  \bibinfo {author} {\bibfnamefont {C.~W.}\ \bibnamefont {Thiel}},\ and\
  \bibinfo {author} {\bibfnamefont {R.~L.}\ \bibnamefont {Cone}},\ }\bibfield
  {title} {\bibinfo {title} {Magnetic g tensors for the $^{4}i_{15/2}$ and
  $^{4}i_{13/2}$ states of er$^{3+}$:\yso{}},\ }\href
  {https://doi.org/10.1103/PhysRevB.77.085124} {\bibfield  {journal} {\bibinfo
  {journal} {Physical Review B - Condensed Matter and Materials Physics}\
  }\textbf {\bibinfo {volume} {77}},\ \bibinfo {pages} {1} (\bibinfo {year}
  {2008})}\BibitemShut {NoStop}%
\bibitem [{\citenamefont {Welinski}\ \emph {et~al.}(2016)\citenamefont
  {Welinski}, \citenamefont {Ferrier}, \citenamefont {Afzelius},\ and\
  \citenamefont {Goldner}}]{Welinski2016}%
  \BibitemOpen
  \bibfield  {author} {\bibinfo {author} {\bibfnamefont {S.}~\bibnamefont
  {Welinski}}, \bibinfo {author} {\bibfnamefont {A.}~\bibnamefont {Ferrier}},
  \bibinfo {author} {\bibfnamefont {M.}~\bibnamefont {Afzelius}},\ and\
  \bibinfo {author} {\bibfnamefont {P.}~\bibnamefont {Goldner}},\ }\bibfield
  {title} {\bibinfo {title} {{High-resolution optical spectroscopy and magnetic
  properties of Yb$^{3+}$ in \yso{}}},\ }\href
  {https://doi.org/10.1103/PhysRevB.94.155116} {\bibfield  {journal} {\bibinfo
  {journal} {Physical Review B}\ }\textbf {\bibinfo {volume} {94}},\ \bibinfo
  {pages} {155116} (\bibinfo {year} {2016})}\BibitemShut {NoStop}%
\bibitem [{\citenamefont {Car}\ \emph {et~al.}(2019)\citenamefont {Car},
  \citenamefont {Veissier}, \citenamefont {Louchet-Chauvet}, \citenamefont
  {Le~Gou\"et},\ and\ \citenamefont {Chaneli\`ere}}]{PhysRevB.100.165107}%
  \BibitemOpen
  \bibfield  {author} {\bibinfo {author} {\bibfnamefont {B.}~\bibnamefont
  {Car}}, \bibinfo {author} {\bibfnamefont {L.}~\bibnamefont {Veissier}},
  \bibinfo {author} {\bibfnamefont {A.}~\bibnamefont {Louchet-Chauvet}},
  \bibinfo {author} {\bibfnamefont {J.-L.}\ \bibnamefont {Le~Gou\"et}},\ and\
  \bibinfo {author} {\bibfnamefont {T.}~\bibnamefont {Chaneli\`ere}},\
  }\bibfield  {title} {\bibinfo {title} {Optical study of the anisotropic
  erbium spin flip-flop dynamics},\ }\href
  {https://doi.org/10.1103/PhysRevB.100.165107} {\bibfield  {journal} {\bibinfo
   {journal} {Phys. Rev. B}\ }\textbf {\bibinfo {volume} {100}},\ \bibinfo
  {pages} {165107} (\bibinfo {year} {2019})}\BibitemShut {NoStop}%
\bibitem [{\citenamefont {Stoll}\ and\ \citenamefont
  {Schweiger}(2006)}]{Stoll2006}%
  \BibitemOpen
  \bibfield  {author} {\bibinfo {author} {\bibfnamefont {S.}~\bibnamefont
  {Stoll}}\ and\ \bibinfo {author} {\bibfnamefont {A.}~\bibnamefont
  {Schweiger}},\ }\bibfield  {title} {\bibinfo {title} {{EasySpin, a
  comprehensive software package for spectral simulation and analysis in
  EPR}},\ }\href {https://doi.org/10.1016/j.jmr.2005.08.013} {\bibfield
  {journal} {\bibinfo  {journal} {Journal of Magnetic Resonance}\ }\textbf
  {\bibinfo {volume} {178}},\ \bibinfo {pages} {42} (\bibinfo {year}
  {2006})}\BibitemShut {NoStop}%
\bibitem [{\citenamefont {Šimėnas}\ \emph {et~al.}(2021)\citenamefont
  {Šimėnas}, \citenamefont {O'Sullivan}, \citenamefont {Kennedy},
  \citenamefont {Lin}, \citenamefont {Fearn}, \citenamefont {Zollitsch},
  \citenamefont {Dold}, \citenamefont {Schmitt}, \citenamefont {Schüffelgen},
  \citenamefont {Liu},\ and\ \citenamefont {Morton}}]{simenas_spin_2021}%
  \BibitemOpen
  \bibfield  {author} {\bibinfo {author} {\bibfnamefont {M.}~\bibnamefont
  {Šimėnas}}, \bibinfo {author} {\bibfnamefont {J.}~\bibnamefont
  {O'Sullivan}}, \bibinfo {author} {\bibfnamefont {O.~W.}\ \bibnamefont
  {Kennedy}}, \bibinfo {author} {\bibfnamefont {S.}~\bibnamefont {Lin}},
  \bibinfo {author} {\bibfnamefont {S.}~\bibnamefont {Fearn}}, \bibinfo
  {author} {\bibfnamefont {C.~W.}\ \bibnamefont {Zollitsch}}, \bibinfo {author}
  {\bibfnamefont {G.}~\bibnamefont {Dold}}, \bibinfo {author} {\bibfnamefont
  {T.}~\bibnamefont {Schmitt}}, \bibinfo {author} {\bibfnamefont
  {P.}~\bibnamefont {Schüffelgen}}, \bibinfo {author} {\bibfnamefont {R.-B.}\
  \bibnamefont {Liu}},\ and\ \bibinfo {author} {\bibfnamefont {J.~J.~L.}\
  \bibnamefont {Morton}},\ }\href {http://arxiv.org/abs/2108.07654} {\bibinfo
  {title} {Spin coherence of near-surface ionised
  \${\textasciicircum}\{125\}\${Te}\${\textasciicircum}+\$ donors in silicon}}
  (\bibinfo {year} {2021}),\ \bibinfo {note} {arXiv:2108.07654
  [cond-mat]}\BibitemShut {NoStop}%
\bibitem [{\citenamefont {Ma}\ \emph {et~al.}(2015)\citenamefont {Ma},
  \citenamefont {Wolfowicz}, \citenamefont {Li}, \citenamefont {Morton},\ and\
  \citenamefont {Liu}}]{PhysRevB.92.161403}%
  \BibitemOpen
  \bibfield  {author} {\bibinfo {author} {\bibfnamefont {W.-L.}\ \bibnamefont
  {Ma}}, \bibinfo {author} {\bibfnamefont {G.}~\bibnamefont {Wolfowicz}},
  \bibinfo {author} {\bibfnamefont {S.-S.}\ \bibnamefont {Li}}, \bibinfo
  {author} {\bibfnamefont {J.~J.~L.}\ \bibnamefont {Morton}},\ and\ \bibinfo
  {author} {\bibfnamefont {R.-B.}\ \bibnamefont {Liu}},\ }\bibfield  {title}
  {\bibinfo {title} {Classical nature of nuclear spin noise near clock
  transitions of bi donors in silicon},\ }\href
  {https://doi.org/10.1103/PhysRevB.92.161403} {\bibfield  {journal} {\bibinfo
  {journal} {Phys. Rev. B}\ }\textbf {\bibinfo {volume} {92}},\ \bibinfo
  {pages} {161403} (\bibinfo {year} {2015})}\BibitemShut {NoStop}%
\bibitem [{\citenamefont {Stevenson}\ \emph {et~al.}(2021)\citenamefont
  {Stevenson}, \citenamefont {Phenicie}, \citenamefont {Gray}, \citenamefont
  {Horvath}, \citenamefont {Welinski}, \citenamefont {Ferrenti}, \citenamefont
  {Ferrier}, \citenamefont {Goldner}, \citenamefont {Das}, \citenamefont
  {Ramesh}, \citenamefont {Cava}, \citenamefont {de~Leon},\ and\ \citenamefont
  {Thompson}}]{https://doi.org/10.48550/arxiv.2110.04876}%
  \BibitemOpen
  \bibfield  {author} {\bibinfo {author} {\bibfnamefont {P.}~\bibnamefont
  {Stevenson}}, \bibinfo {author} {\bibfnamefont {C.~M.}\ \bibnamefont
  {Phenicie}}, \bibinfo {author} {\bibfnamefont {I.}~\bibnamefont {Gray}},
  \bibinfo {author} {\bibfnamefont {S.~P.}\ \bibnamefont {Horvath}}, \bibinfo
  {author} {\bibfnamefont {S.}~\bibnamefont {Welinski}}, \bibinfo {author}
  {\bibfnamefont {A.~M.}\ \bibnamefont {Ferrenti}}, \bibinfo {author}
  {\bibfnamefont {A.}~\bibnamefont {Ferrier}}, \bibinfo {author} {\bibfnamefont
  {P.}~\bibnamefont {Goldner}}, \bibinfo {author} {\bibfnamefont
  {S.}~\bibnamefont {Das}}, \bibinfo {author} {\bibfnamefont {R.}~\bibnamefont
  {Ramesh}}, \bibinfo {author} {\bibfnamefont {R.~J.}\ \bibnamefont {Cava}},
  \bibinfo {author} {\bibfnamefont {N.~P.}\ \bibnamefont {de~Leon}},\ and\
  \bibinfo {author} {\bibfnamefont {J.~D.}\ \bibnamefont {Thompson}},\ }\href
  {https://doi.org/10.48550/ARXIV.2110.04876} {\bibinfo {title}
  {Erbium-implanted materials for quantum communication applications}}
  (\bibinfo {year} {2021})\BibitemShut {NoStop}%
\bibitem [{\citenamefont {Kanai}\ \emph {et~al.}(2022)\citenamefont {Kanai},
  \citenamefont {Heremans}, \citenamefont {Seo}, \citenamefont {Wolfowicz},
  \citenamefont {Anderson}, \citenamefont {Sullivan}, \citenamefont {Onizhuk},
  \citenamefont {Galli}, \citenamefont {Awschalom},\ and\ \citenamefont
  {Ohno}}]{doi:10.1073/pnas.2121808119}%
  \BibitemOpen
  \bibfield  {author} {\bibinfo {author} {\bibfnamefont {S.}~\bibnamefont
  {Kanai}}, \bibinfo {author} {\bibfnamefont {F.~J.}\ \bibnamefont {Heremans}},
  \bibinfo {author} {\bibfnamefont {H.}~\bibnamefont {Seo}}, \bibinfo {author}
  {\bibfnamefont {G.}~\bibnamefont {Wolfowicz}}, \bibinfo {author}
  {\bibfnamefont {C.~P.}\ \bibnamefont {Anderson}}, \bibinfo {author}
  {\bibfnamefont {S.~E.}\ \bibnamefont {Sullivan}}, \bibinfo {author}
  {\bibfnamefont {M.}~\bibnamefont {Onizhuk}}, \bibinfo {author} {\bibfnamefont
  {G.}~\bibnamefont {Galli}}, \bibinfo {author} {\bibfnamefont {D.~D.}\
  \bibnamefont {Awschalom}},\ and\ \bibinfo {author} {\bibfnamefont
  {H.}~\bibnamefont {Ohno}},\ }\bibfield  {title} {\bibinfo {title}
  {Generalized scaling of spin qubit coherence in over 12,000 host materials},\
  }\bibfield  {journal} {\bibinfo  {journal} {Proceedings of the National
  Academy of Sciences}\ }\textbf {\bibinfo {volume} {119}},\ \href
  {https://doi.org/10.1073/pnas.2121808119} {10.1073/pnas.2121808119} (\bibinfo
  {year} {2022})\BibitemShut {NoStop}%
\end{thebibliography}%

\end{document}


\beginsupplement

\title{Coherent spin dynamics of rare-earth doped crystals in the high-cooperativity regime: Supplementary material}

\author{Joseph~Alexander}
\email[]{joseph.alexander.18@ucl.ac.uk}
\affiliation{London Centre for Nanotechnology, University College London, London WC1H 0AH, United Kingdom}

\author{Gavin~Dold}

\affiliation{London Centre for Nanotechnology, University College London, London WC1H 0AH, United Kingdom}
\affiliation{National Physical Laboratory, Hampton Road, Teddington TW11 0LW, United Kingdom}

\author{Oscar~W.~Kennedy}
\affiliation{London Centre for Nanotechnology, University College London, London WC1H 0AH, United Kingdom}

\author{Mantas~\v{S}im\.{e}nas}
\affiliation{London Centre for Nanotechnology, University College London, London WC1H 0AH, United Kingdom}

\author{James~O'Sullivan}
\affiliation{London Centre for Nanotechnology, University College London, London WC1H 0AH, United Kingdom}

\author{Christoph~W.~Zollitsch}
\affiliation{London Centre for Nanotechnology, University College London, London WC1H 0AH, United Kingdom}

\author{Sacha Welinski}
\affiliation{Universit\'e PSL, Chimie ParisTech, CNRS, Institut de Recherche de Chimie Paris, 75005 Paris, France}
\author{Alban Ferrier}
\affiliation{Universit\'e PSL, Chimie ParisTech, CNRS, Institut de Recherche de Chimie Paris, 75005 Paris, France}
\affiliation{Facult\'e des Sciences et Ing\'enierie,  Sorbonne Universit\'e, 75005 Paris, France}
\author{Eloïse Lafitte-Houssat}
\affiliation{Universit\'e PSL, Chimie ParisTech, CNRS, Institut de Recherche de Chimie Paris, 75005 Paris, France}
\affiliation{Thales Research and Technology, 1 Avenue Augustin Fresnel, 91767 Palaiseau, France}

\author{Philippe Goldner}
\affiliation{Universit\'e PSL, Chimie ParisTech, CNRS, Institut de Recherche de Chimie Paris, 75005 Paris, France}
\author{Tobias Lindstr\"om}
\affiliation{National Physical Laboratory, Hampton Road, Teddington TW11 0LW, United Kingdom}
\author{John J. L. Morton}
\email[]{jjl.morton@ucl.ac.uk}
\affiliation{London Centre for Nanotechnology, University College London, London WC1H 0AH, United Kingdom}
\affiliation{Department of Electronic and Electrical Engineering, UCL, London WC1E 7JE, United Kingdom}

\date{\today}

\maketitle

\section{Experimental setup}

The experiments in this paper were performed within a dilution refrigerator (BlueFors LD400) with a magnetic field supplied via a (1,1,3)T vector magnet. The superconducting resonators consisted of planar NbN, patterned into a `thin-ring' design \cite{Dold2019,doi:10.1063/1.5129032} for the \nd{145} sample (spiral design \cite{PhysRevApplied.12.024021} for \yb{}) using photolithography. The samples were mounted in the puck differently, the \nd{145} sample was encased within a 3D cavity to suppress radiative emission to the environment, and measured in transmission via two weakly coupled antennae. The \yb{} samples were placed on top of a PCB consisting of a co-planar waveguide with a constriction in the centre. The sample is placed after the constriction so the emitted signal is prevented from returning up the input path by being reflected and travelling up the output path.

\begin{figure}[h]
	\includegraphics[width = 0.50\columnwidth]{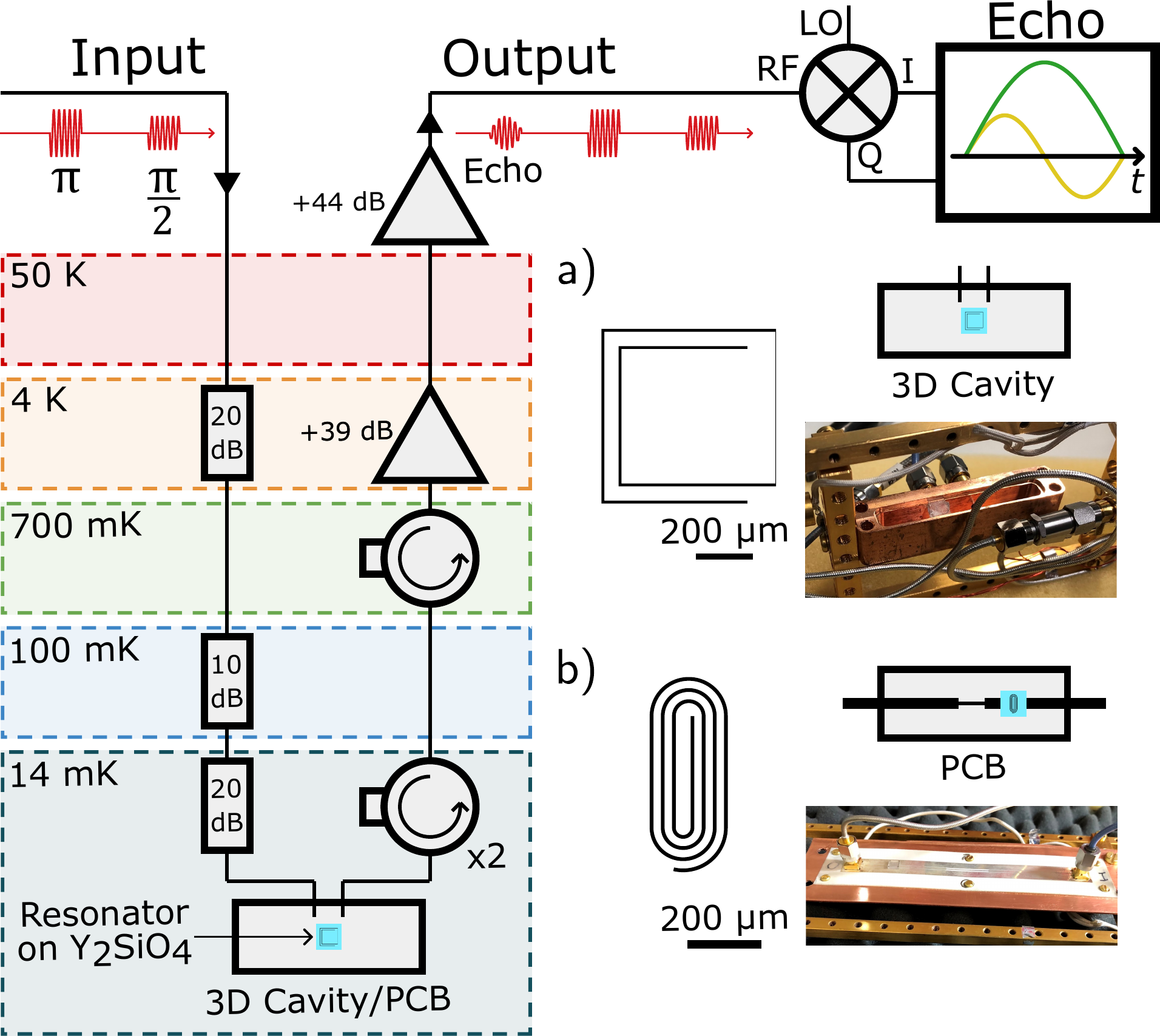}%
	\caption{Experimental schematic, showing the devices installed within a dilution refrigerator equipped with coaxial cables to measure the device in transmission. Right: schematics of (a) the thin-ring encased in a cavity and (b) spiral designs placed on a PCB, which were fabricated on \SI{200}{ppm} \nd{145}:\yso{} and \SI{50}{ppm} (\SI{5}{ppm}) \yb{nat(171)}:\yso{} respectively.}
	\label{fig:setup}
\end{figure}

The overall measurement setup is shown schematically in Fig.~\ref{fig:setup}. Coaxial cables carry signals to the 3D cavity (PBC) within the fridge, attenuated by 50 dB to reduce thermal noise at the sample. The returned signal is amplified with a cryogenic low-noise amplifier and a room-temperature amplifier. Continuous-wave measurements were performed by connecting a vector network analyser across the input and output of the fridge. The spin dynamics were investigated by performing pulsed spectroscopy using a custom-built ESR spectrometer. Pulses at \si{GHz} frequencies are generated by modulating a vector signal generator (VSG) using an arbitrary waveform generator, and sent down the input line. The returned signal from the fridge then passes through a fast RF switch which protects the spectrometer from high power pulses, and is demodulated with an I/Q mixer referenced against the local oscillator of the VSG. 

\section{Rare-Earth Spin Hamiltonian}

The Nd and Yb impurities in YSO are described by the spin Hamiltonian:
\begin{equation}
    \mathcal{H} = \mathbf{S}\cdot \hat{A} \cdot \mathbf{I} +\mu_{B} \mathbf{B} \cdot \hat{g} \cdot \mathbf{S} - \mu_{n}\mathbf{B} \cdot g_{n} \cdot \mathbf{I}
\end{equation}

where the electron spin ($\mathbf{S}$) interacts with the nuclear spin ($\mathbf{I}$) described by a hyperfine tensor, $\hat{A}$, and to an external magnetic field ($\mathbf{B}$) via a g-tensor, $\hat{g}$. The nuclear spin also interacts with the field via a nuclear g-factor ($g_n$). In this work the nuclear spin transitions are ignored. The g-tensor and hyperfine (A) tensor for each species are outlined below:

\textbf{\nd{145}} \cite{H_Maier_Flaig}:

\begin{equation}
   \hat{g}_{Nd}  = \begin{pmatrix}
    1.30 && 0.62 && 0.22 \\
    0.62 && 0.22 && -2.07 \\
    1.62 && 1.62 && -2.86 
\end{pmatrix}_{(D1,D2,b)}
\end{equation}
\begin{equation}
   \hat{A}_{Nd}  = \begin{pmatrix}
    -37.1 && -99.9 && -83.4 \\
    -99.9 && -589.2 && 169.4 \\
    -83.4 && 169.4 && -678.4 
\end{pmatrix}_{(D1,D2,b)}
\end{equation}

\textbf{\yb{171} Site 1 \cite{Tiranov2018}}:

\begin{equation}
   \hat{g}_{Yb,1}  = \begin{pmatrix}
    6.072 && -1.46 && -0.271 \\
    -1.460 && 1.845 && -0.415 \\
    -0.271 && -0.415 && 0.523 
\end{pmatrix}_{(D1,D2,b)}
\end{equation}
\begin{equation}
   \hat{A}_{Yb,1}  = \begin{pmatrix}
    4.847 && -1.232 && -0.244 \\
    -1.232 && 1.425 && -0.203 \\
    -0.244 && -0.203 && 0.618 
\end{pmatrix}_{(D1,D2,b)} \times 10^{3}
\end{equation}

\textbf{\yb{171} Site 2 \cite{Tiranov2018}}:

\begin{equation}
   \hat{g}_{Yb,2}  = \begin{pmatrix}
    0.999 && -0.766 && -0.825 \\
    -I 0.766 && 0.825 && -0.424 \\
    0.825 && -0.424 && 5.867 
\end{pmatrix}_{(D1,D2,b)}
\end{equation}
\begin{equation}
   \hat{A}_{Yb,2}  = \begin{pmatrix}
    0.686 && -0.718 && 0.492 \\
    -0.718 && 0.509 && -0.496 \\
    0.492 && -0.496 && 4.729 
\end{pmatrix}_{(D1,D2,b)} \times 10^{3}
\end{equation}

\section{Background signal in echo-detected avoided crossing}

\begin{figure}[h]
    \centering
    \includegraphics[width=0.5\columnwidth]{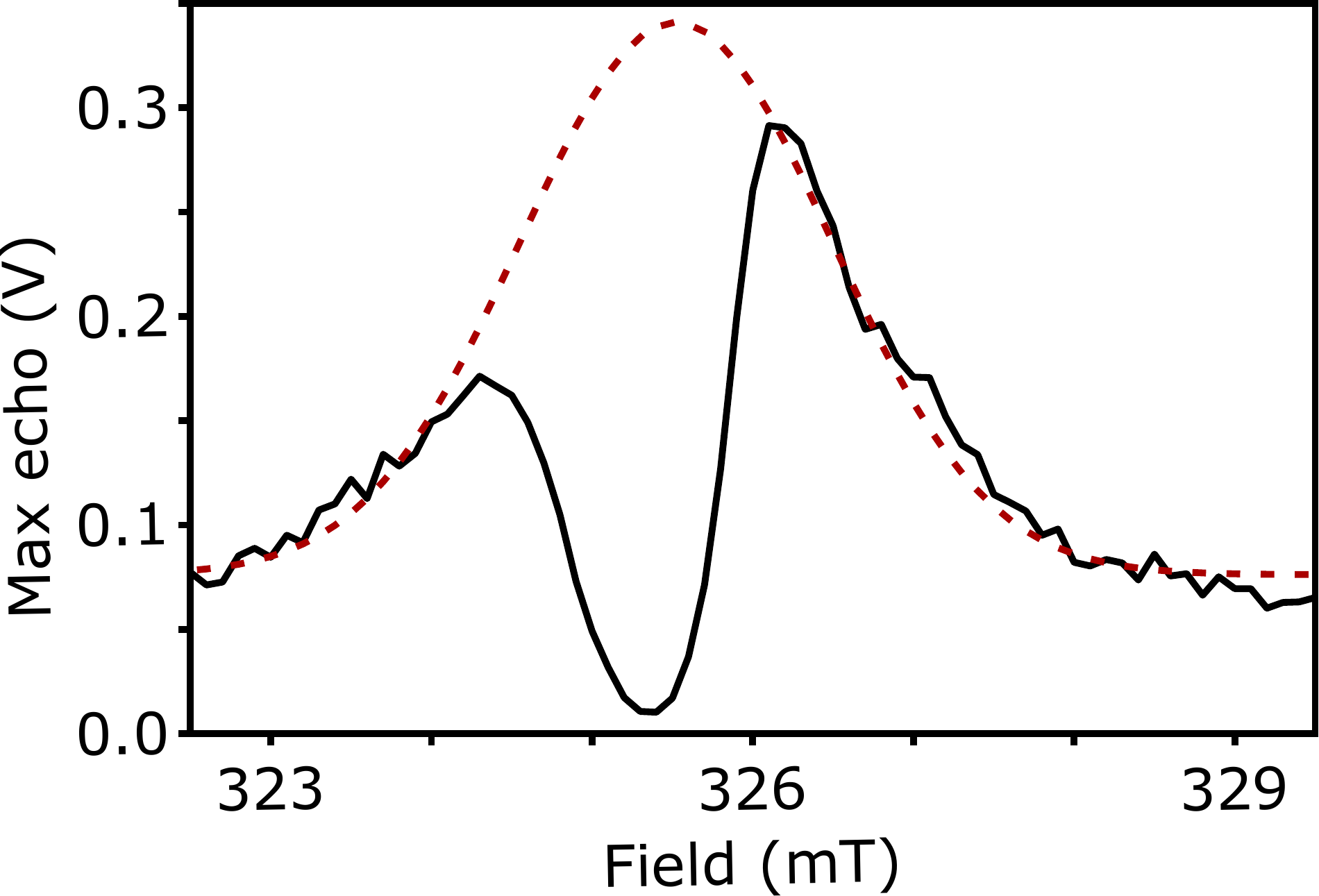}
    \caption{Maximum echo at each field step of the echo-detected avoided crossing (Fig. 1c main text). Fitted to this (dashed red), is a Gaussian with linewidth \SI{0.97}{\milli\tesla}, a background echo is observed at the edges of this fit with an intensity of \SI{0.76}{\volt}.}
    \label{fig:background_echo}
\end{figure}

In Fig. 1c of the main text a signal is measured in a two-pulse Hahn echo well away from the centre of the \nd{145} spin line. To highlight this weak background echo we plot the maximum echo at each field in Fig.~\ref{fig:background_echo}. The echo intensity drops at the centre of the spin line where the resonator and the spins have fully hybridised and we can no longer probe with the resonator. By fitting a Gaussian to the maximum echo using a linewidth equal to that measured using the dispersive frequency shift of the resonator (\SI{3}{\mega\hertz} - Fig.~\ref{fig:cooperativty}a) divided by the gyromagnetic ratio of the transition, we observe a good fit to the data and record a background echo intensity of \SI{0.76}{\volt}. This background echo could be due to other rare-earth ions which may be present or $g=2$ spins. 

\section{Cooperativity}

To determine the cooperativity between the resonator and the spin system the dispersive shift of the resonant frequency and the resonator linewidth are measured as the field is swept across the spin line. As the resonator and the spins hybridise the measured resonator frequency ($f$) follows that of the spin line and its quality factor decreases. This can be expressed as \cite{doi:10.1063/1.3601930}:

\begin{align}
    f = f_{0} - \frac{g_{ens}^{2}\Delta}{\Delta^{2}+\gamma^{2}}\\
    \kappa = \kappa_{0} + \frac{g_{ens}^{2}\gamma}{\Delta^{2}+\gamma^{2}},
\end{align}\label{eqn:coop}

where $\kappa$ is the resonator half-width, $g_{ens}$ is the ensemble coupling strength, $\gamma$ is the spin half-linewidth and $\Delta$ is the detuning of the field from resonance given by $df/dB \cdot (B-B_{0})$ where $B_{0}$ is the field at which the spins and resonator are resonant.

\begin{figure}[h]
    \centering
    \includegraphics[width=0.50\columnwidth]{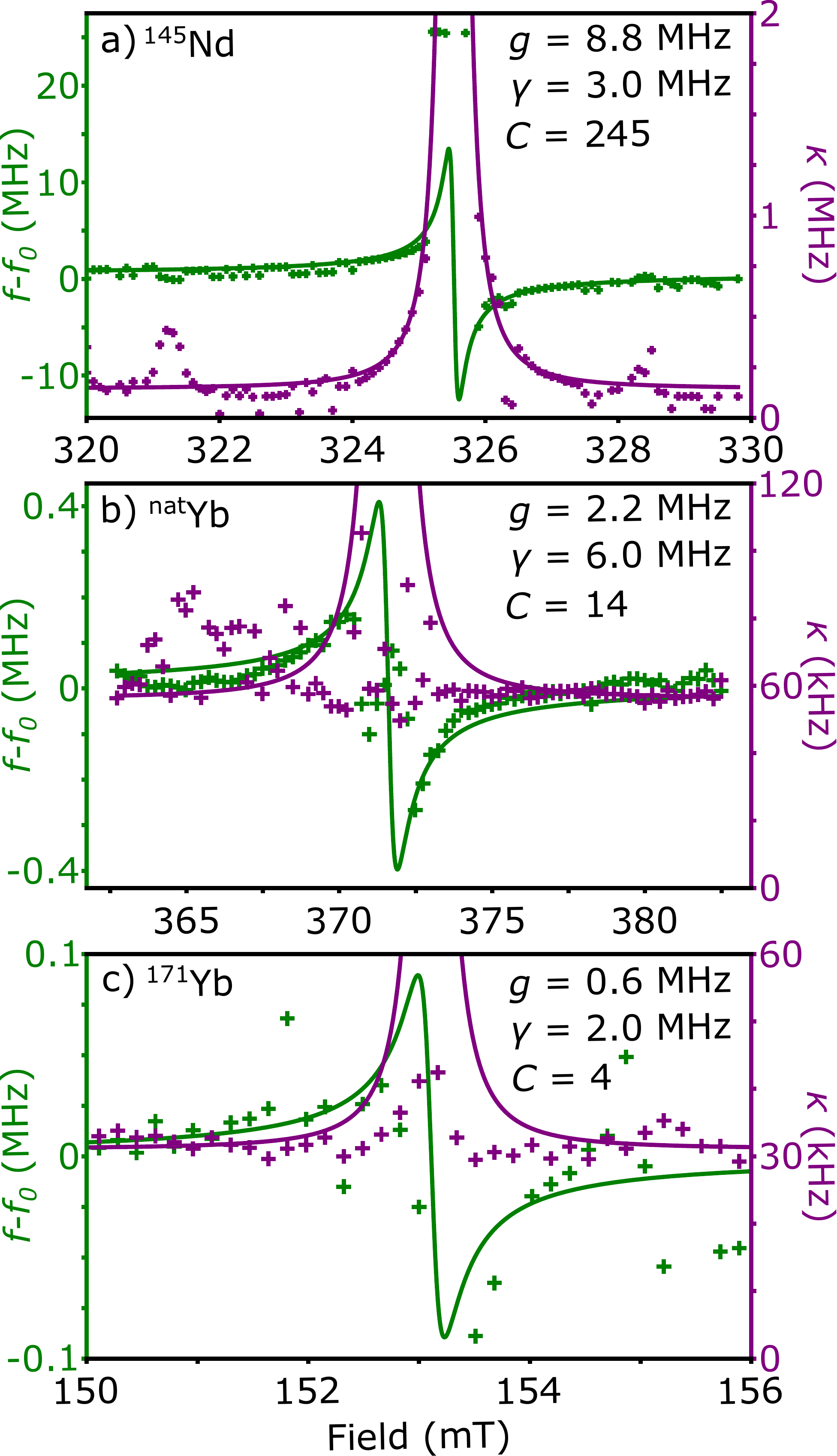}
    \caption{Resonator frequency shift and linewidth broadening for the three different samples. Equation \ref{eqn:coop}} is used to extract the ensemble coupling strength, spin linewidth and the cooperativity.
    \label{fig:cooperativty}
\end{figure}

Fig.~\ref{fig:cooperativty} shows the dispersive frequency shifts and resonator linewidth broadening for transitions in the three different systems. The extracted parameters are highlighted with cooperatives (C = $g^{2}/\kappa\gamma$) of 245, 14 and 4 measured for \nd{145}, \yb{nat} and \yb{171} respectively.

\section{Instantaneous Diffusion Model}

The effect of instantaneous diffusion (ID) on the coherence time (T$_{2}$) can be described by \cite{Tyryshkin2011a}:

\begin{align}\label{eqn:T2_ID}
    T_{2} = \frac{8h}{5\mu_{0}(g\mu_{\mathrm{B}})^{2}n}\mathrm{sin}^{2}\left ( \frac{\theta}{2} \right ),
\end{align}

where $g$ is the effective g-factor, $n$ is the excited spin density and $h$, $\mu_{0}$ and $\mu_{\mathrm{B}}$ are Plank's constant, vacuum permeability and the Bohr magneton respectively. $\theta$ is the angle of the second pulse in the Hahn echo sequence, this is $\pi$ in this experiment meaning that the effect of ID is maximal. 

\begin{figure}
    \centering
    \includegraphics[width=0.5\columnwidth]{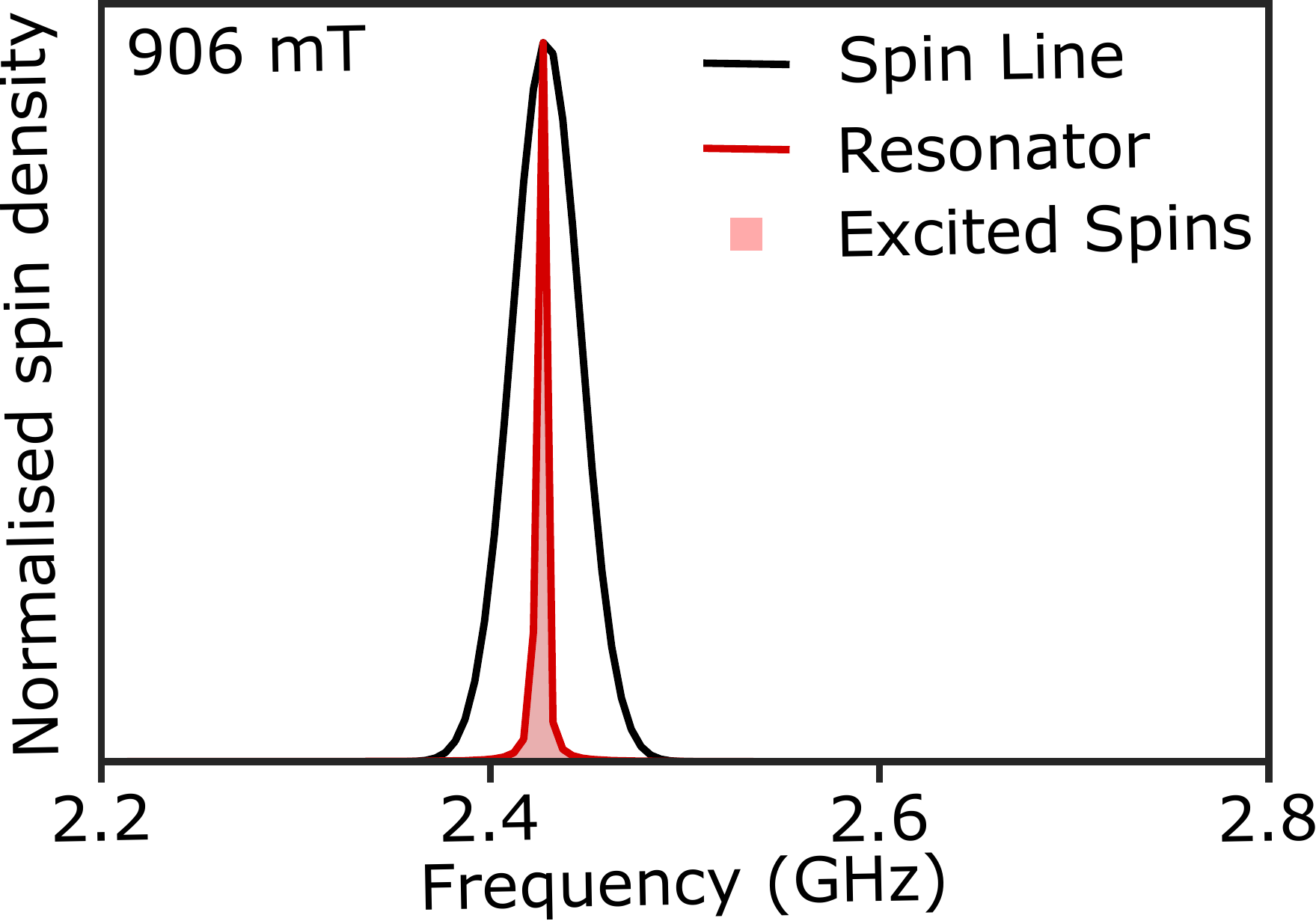}
    \caption{The interaction between the spin line and the resonator during a field sweep. The number of excited spins used in equation \ref{eqn:T2_ID} is the proportion of the spin line within the resonator bandwidth, shown by the red shaded region. As the field moves off and on resonance the spin line moves over the resonator and so the excited spin density reaches a maximum when both are resonant with each-other, shown here at \SI{906}{\milli\tesla}.}
    \label{fig:spin_res_interaction}
\end{figure}

The excited spin density is not constant during the field sweep. To approximate this the total spin density is modelled as a Gaussian with linewidth $\gamma = \SI{8.7}{\mega\hertz}$ -- calculated from fitting to the dispersive frequency shift -- the linewidth of the resonator is narrower than the spins and therefore the spin density is taken as the proportion of the spins which are within the resonator bandwidth described by a Lorentzian. An excitation $\pi/2$ pulse of \SI{7.5}{\micro\second} acts as a filter on the resonator limiting its linewidth to \SI{133}{\kilo\hertz}. FIG. \ref{fig:spin_res_interaction} shows the fraction of spins which interact with the resonator at a single field at the centre of the avaoided crossing. Eqn. \ref{eqn:T2_ID} is then used to determine the limit on T$_{2}$ as a function of field.

\section{Accounting for the noise floor}

When measuring an echo amplitude, the voltage noise does not time-average to zero as it would for I or Q separately, but instead the noise power ($\propto V^2$) adds to the signal power. This prevents the measured signal from decaying to zero at long timescales. The appropriate way to describe to such an amplitude decay measured in voltage is
\begin{equation*}
\textrm{amplitude} = \sqrt{ \textrm{signal}^2 + \textrm{noise}^2 }.\label{eqn:supp:noise_desc}
\end{equation*}

Hence the equation used to fit to the single exponential decay as a function of pulse separation time $\tau$ in a two-pulse $T_2$ measurement is
\begin{equation}
A(\tau) = \sqrt{ \left( A_0 e^{-\frac{2\tau}{T_2}} \right)^2 + C^2 }\label{eqn:supp:noise_t2}
\end{equation}
where $A_0$ is the magnitude of the exponential decay, and $C$ is the noise floor which can be determined by measuring the retrieved signal in the absence of any applied pulses.

\section{Relaxation time $T_1$}

\begin{figure}
	\includegraphics[width = 0.5\columnwidth]{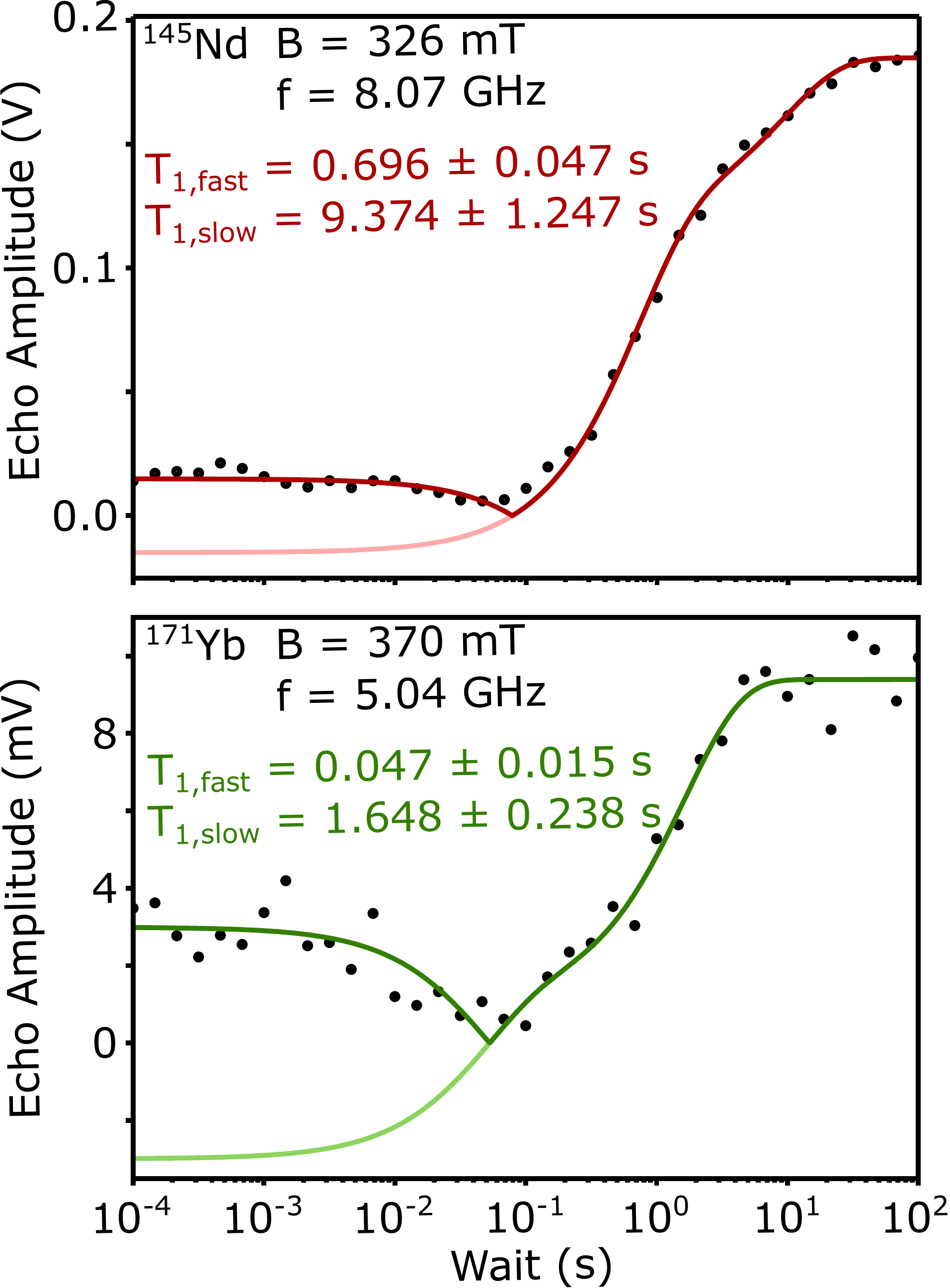}%
	\caption{Measurement of $T_1$ at \SI{14}{mK} by inversion recovery in both systems. The data represent echo amplitudes, resulting in all echoes having positive amplitude. A biexponential fit (dark line) to the data approximates the continuum of relaxation rates due to an inhomogeneously Purcell-enhanced spin ensemble with two characterstic rates. The lighter line indicates the region where the echo is inverted, which is otherwise hidden due to measuring the echo amplitude.}
	\label{fig:T1}
\end{figure}

In order to accurately model and interpret decoherence data we must have a measure of the spin relaxation, \textit{T$_{1}$}. In systems where the spins are strongly coupled to the resonator, two relaxation mechanisms come into play: spin--lattice relaxation \cite{Cruzeiro2016,Lim2017}, and Purcell-enhanced relaxation whereby a spin coupled to a cavity with strength $g_0$ relaxes via emission into a resonant cavity with characteristic time $T_1 = \tfrac{\kappa_\mathrm{s}}{4 g_0^2}$ \cite{RANJAN2020106662}. Since $g_0$ is proportional to the $B_1$ field amplitude, which for this resonator decays as $\tfrac{1}{r}$ with distance $r$ from the inductor, we expect a continuum of relaxation rates, ranging from fast relaxation by strongly-coupled spins near the resonator to slower bulk-like relaxation due to spins further away.

We use an inversion-recovery method, using a WURST (Wideband, Uniform Rate, Smooth Truncation) pulse \cite{ODELL201328} to adiabatically invert the measured spin ensemble, allowing relaxation processes to proceed for wait time $T_\mathrm{w}$, followed by a two-pulse sequence to produce an echo (WURST--$T_\mathrm{w}$--$\tfrac{\pi}{2}$--$\tau$--$\pi$--$\tau$--echo). A WURST pulse \SI{800}{\micro\second} in length and chirped by \SI{2}{MHz} achieves the desired inversion.

Varying $T_\mathrm{w}$ from \SI{100}{s} to \SI{100}{\micro\second} at \SI{14}{mK} produces the data presented in Fig.~\ref{fig:T1}. As expected, the data deviate significantly from a single exponential. As an approximation to the wide range of $T_1$ expected from an inhomogeneously Purcell-enhanced spin ensemble, we fit a biexponential curve to extract two characteristic rates $T_{1,\mathrm{fast}}$ and $T_{1,\mathrm{slow}}$ where, within the approximation described above, $T_{1,\mathrm{fast}}$ describes a rate at which strongly coupled spins are relaxing via their coupling to the resonator, and $T_{1,\mathrm{slow}}$ is a contribution from a sub-ensemble more weakly coupled spins.

\section{Covariance of R and $\Gamma_{SD}$}

\begin{figure}[h]
    \centering
    \includegraphics[width=0.75\columnwidth]{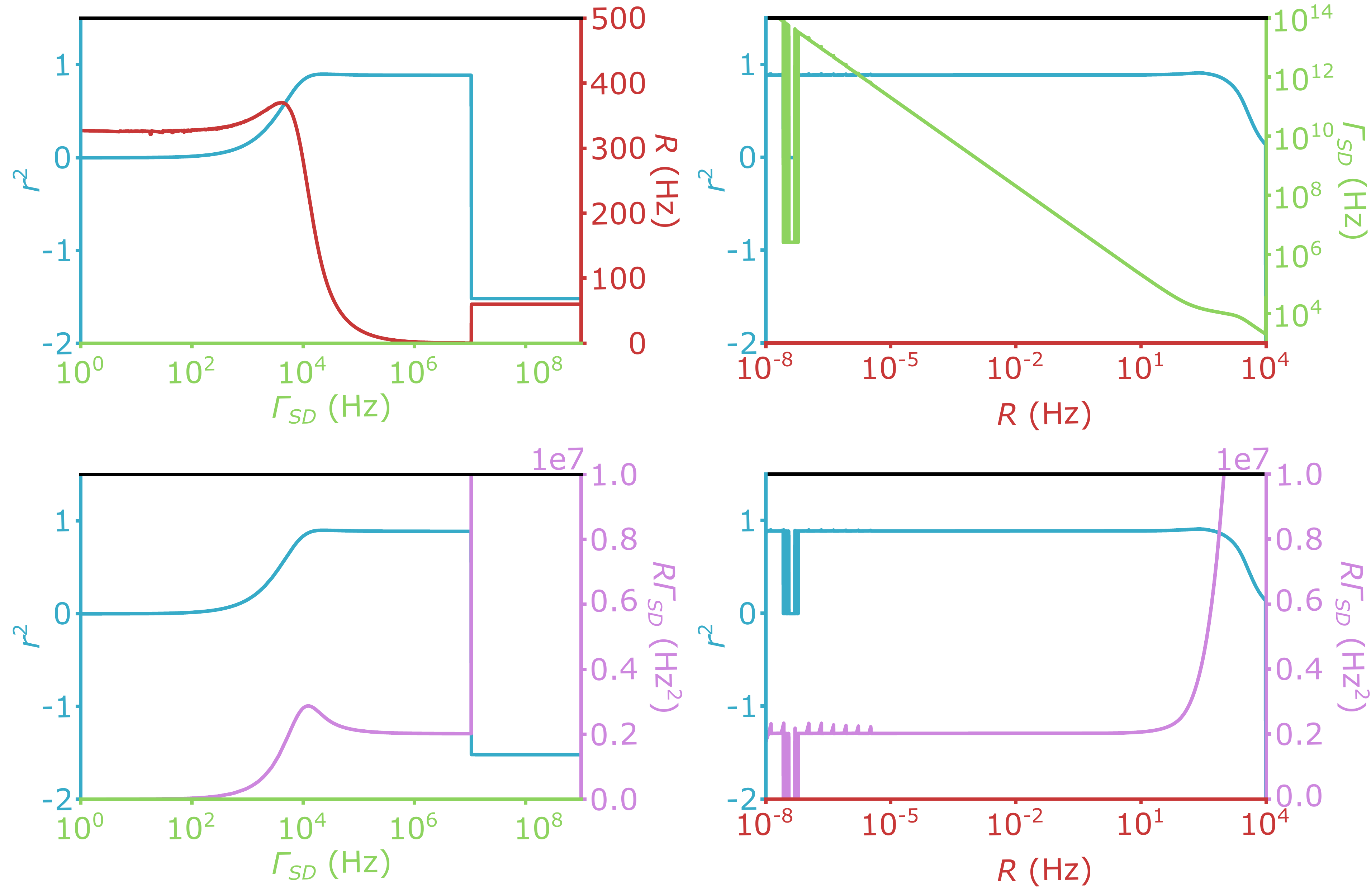}
    \caption{Covariance of $\Gamma_{SD}$ amd R, while fitting to the three pulse echo data, $\Gamma_{SD}$ (R) are fixed while the other is fitted, the resulting r$^{2}$ value is plotted. It can be seen that there is a range of $\Gamma_{SD}$ (R) values that result in a high r$^{2}$ thus meaning $\Gamma_{SD}$ and R cannot be uniquely identified. The product of the two is also plotted, this is constant across the region of high r$^{2}$ and thus is the appropriate quantity to use in the analysis.}
    \label{fig:covariance}
\end{figure}

Three parameters are needed to fit to three pulse stimulated echo data, these are the flip-flop rate (R), the spectral diffusion linewidth ($\Gamma_{SD}$) and the residual decoherence ($\Gamma_{0}$). It was observed that R and $\Gamma_{SD}$ could not be uniquely determined due to their covariance. This is shown by figure \ref{fig:covariance}, here we fix both $\Gamma_{SD}$ and $R$ in turn and measure the $r^{2}$ value of the resulting fit. What results is a range of $\Gamma_{SD}$/$R$ values give $r^{2}$ values close to 1. This means the fitting routine cannot uniquely determine $\Gamma_{SD}$ or $R$, however when we plot the product of these two, it can be seen that $R\Gamma_{SD}$ is constant over the region with high $r^{2}$. This means to accurately interpret the three pulse echo data we should use $R\Gamma_{SD}$ as the fitted parameter. Using analysis in the main text, this allows for a spectral diffusion limited T$_{2}$ to be extracted. 

\section{Thermalisation of the $^{145}$Nd sample}

\begin{figure}[ht]
    \centering
    \includegraphics[width=0.5\columnwidth]{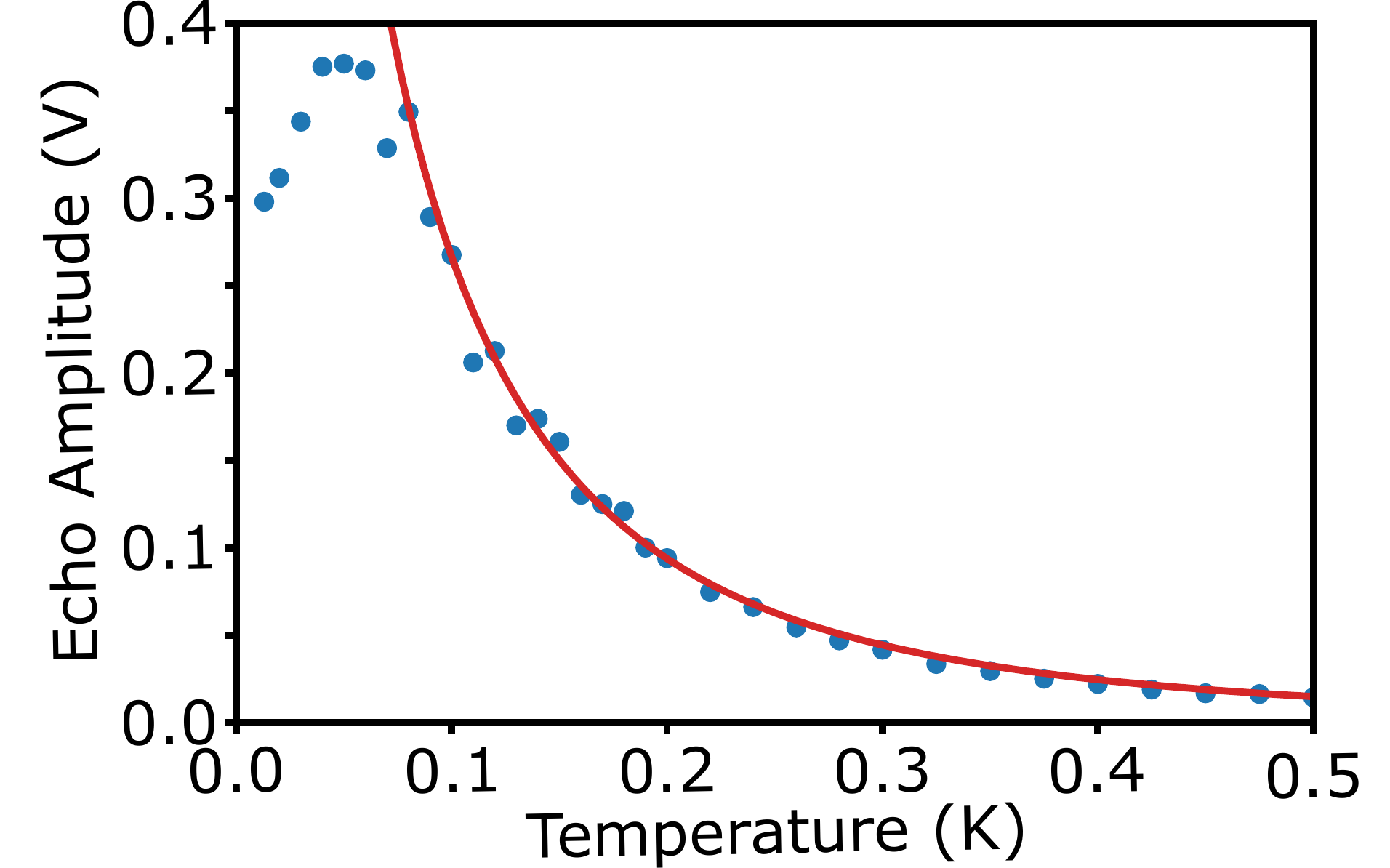}
    \caption{Echo amplitude of the first data point in the T$_{2}$ temperature dependence in \nd{145} measurement. Plotted in red is the polarisation of the measured transition, which is maximal at \SI{0}{\kelvin}. The poor themalisiation can be seen where the echo amplitude deviates from the polarisation at approximately \SI{100}{\milli\kelvin}.}
    \label{fig:my_label}
\end{figure}

The temperature dependence of the coherence time in \nd{145} returns a large residual decoherence rate, even at the recorded base temperature of the dilution fridge. This is believed to be in part to poor thermalisation between the mixing chamber plate and the sample. Evidence for this occurs when looking at the echo amplitude as a function of temperature. In this system we expect the echo amplitude to increase as the temperature decreases as we are measuring the $M_{I} = 7/2$ transition and so it should be maximally populated at \SI{0}{\kelvin}. The increase in echo amplitude deviates from the increase in polarisation at approximately \SI{100}{\milli\kelvin}, this corresponds well to the analysis in the main text from the stimulated three pulse echo data. This poor thermalisation may have been due to the use of the 3D cavity, and suggests that longer coherence times could be achieved in \nd{145} if this were resolved.

\section{Spectral Diffusion from \y{89}}

While spectral diffusion from electron spin flips can be suppressed by reducing the temperature of the sample (as shown in the main text), the nuclear spin bath of the crystal can also cause spectral diffusion. \y{89} is $100\%$ abundant with nuclear spin 1/2, this results in a large nuclear spin bath in \yso{} crystals. The rate of nuclear spin flip-flops in \yso{} is very low and so the effect of the nuclear spin bath can only be seen at the longest time scales - we reach these time scales at low temperatures in the \yb{} sample. In \cite{Bottger2006} a `method-of-moments' approach is used to find the effect of \y{89} spectral diffusion on a central spin via the spectral diffusion linewidth:

\begin{equation}
\Gamma_{SD,Y} = 0.14\mu_{0}\gamma_{Y}g\mu_{B}n_{Y}\sqrt{I(I+1)}\mathrm{sech\left(\frac{h\gamma_{Y}B}{2k_{B}T}\right)},
\end{equation}
where $\gamma_{Y}$, $n_{Y}$ and $I$ are the gyromagnetic ratio, spin density and nuclear spin of \y{89} respectively. $g$ is the effective g-factor of the central \yb{} spin in the nuclear spin bath.

Assuming the main \y{89} spectral diffusion arises from nuclear flip-flops within the same site, an upper bound is placed on the rate via \cite{Bottger2006}:

\begin{equation}
    R = 0.25\mu_{0}h\gamma_{Y}^{2}\frac{n_{Y}}{2}\sqrt(I(I+1))\mathrm{sech\left(\frac{h\gamma_{Y}B}{2k_{B}T}\right)}.
\end{equation}

Using this, the linewidth broadening of \yb{171} from \y{89} induced spectral diffusion at \SI{370}{\milli\tesla} is computed to be $\Gamma_{SD,Y} = $\SI{38}{\kilo\hertz} with a nuclear flip-flop rate $R = $\SI{3.6}{\hertz}. This gives $R_{Y}\Gamma_{SD,Y} = \SI{1.05e5}{\hertz\squared}$, comparing this to the measured $R\Gamma_{SD}$ from the stimulated echo measurements ($R\Gamma_{SD} = \SI{1.3+-0.1e5}{\hertz\squared}$) gives good agreement and results in a limiting $T_{2}$ of \SI{3.48}{\milli\second}. This shows that the coherence time of \yb{171} at \SI{5.04}{\giga\hertz} and \SI{370}{\milli\tesla} is limited by nuclear spin flip-flops within the \y{89} nuclear spin bath.

\section{Angular Dependent T$_{2}$}

To model the angular dependence of the \yb{171} coherence time we assume that it is spectral diffusion limited. Using the spectral diffusion rate (\emph{R}) and linewidth (\emph{$\Gamma_{SD}$}) a spectral diffusion limited \emph{T}$_{2}$ can be deduced. With each angle, the working field is calculated using easyspin, this is the resonant field of \yb{171} at \SI{2.43}{\giga\hertz}. After setting the field, the frequencies and thus the Zeeman temperature of all the environmental subensembles can be calculated as well as their effective g-factors. Using this the same analysis as in the temperature dependant fits can used.

The coherence time was modeled using~\cite{Bottger2006,Lim2017,Cruzeiro2016}:
\begin{equation}
    T_2 = \frac{1}{\Gamma} = \frac{2}{R\Gamma_{\mathrm{SD}}} \left ( \sqrt{\Gamma_{0}^{2} + \frac{R\Gamma_{\mathrm{SD}}}{\pi}} - \Gamma_{0} \right ) \approx \frac{2}{\sqrt{\pi R \Gamma_{\rm SD}}},
    \label{eqn:T2SD}
\end{equation}

where $R$ and $\Gamma_{SD}$ are now angular dependant, 

\begin{equation}
    R_i(\theta) = \beta_{\rm ff}\left ( \theta \right ) \frac{n_i^2}{\Gamma_i}\sech^2\Big(\frac{T_{\mathrm{Z},i}}{T}\Big)
    \label{eqn:R}
\end{equation}

and 

\begin{equation}
    \Gamma_{\mathrm{SD},i}(\theta) = \frac{\pi\mu_0\mu_B^2}{9\sqrt{3}h}n_ig_ig\sech^2\Big(\frac{T_{\mathrm{Z},i}}{T}\Big).
    \label{eqn:gammaSD}
\end{equation}

following the same analysis as in the main text the spectral diffusion limited decoherence is modelled as: 

\begin{equation}\label{eqn:t2oft}
\Gamma(\theta) = \Gamma_\mathrm{res} + \sum_i \frac{\xi n_i^{3/2}M_i}{\Big(1+e^{\tfrac{T_\mathrm{Z}^{i}}{T}}\Big)\Big(1+e^{-\tfrac{T_\mathrm{Z}^{i}}{T}}\Big)}.
\end{equation}

By setting the temperature ($T$) to \SI{14}{\milli\kelvin} and the fit parameter $\xi = 6$ from the temperature study, an angular dependant decoherence rate can be calculated, this is plotted in FIG. 4 in the main text.

\section{ZEFOZ in $^{171}$Yb:Y$_{2}$SiO$_{5}$}

\begin{figure}
    \centering
    \includegraphics[width=0.5\columnwidth]{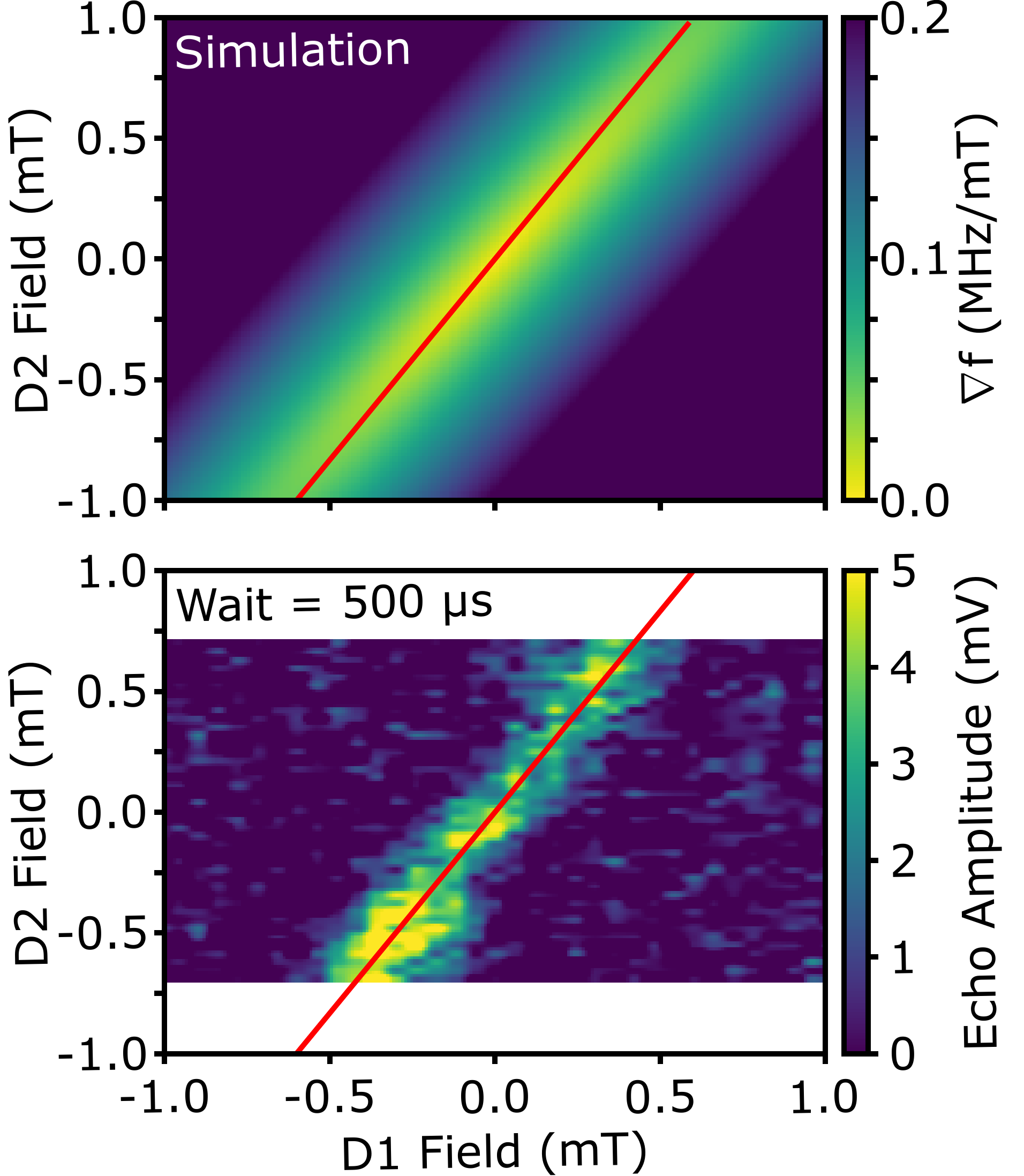}
    \caption{The low $df/dB$ landscape of \yb{171} around zero field. The above panel shows how $\nabla$f varies at low fields in the D1-D2 plane, there is a clear line of low effective g-factor as indicated by the red line at \SI{49}{^{\circ}}. The bottom panel shows the Hahn echo amplitude after a wait of \SI{500}{\micro\second} between pulses. The regions with high echo amplitude indicate regions with long T$_{2}$, this corresponds well with the line of low g-factor from the simulation}
    \label{fig:zefoz heatmap}
\end{figure}

To explore the region of low effective g-factor around zero field in the \yb{171}:\yso{} system we performed a field swept Hahn echo. By using a long wait time of \SI{500}{\micro\second} between the two pulses in a Hahn echo sequence, regions with long \textit{T$_{2}$} can be identified. In this experiment regions with short coherence times will have a much smaller echo as the signal will have significantly decayed. By simulating the effective g-factor with EasySpin a line of low $\nabla_{B} f$ was identified, this matches up well with experiment where this region at $49^{\circ}$ exhibited the largest echo as shown in FIG. \ref{fig:zefoz heatmap}. This line was then chosen to perform the field dependant \textit{T$_{2}$} measurement in the main text.

\begin{figure}
    \centering
    \includegraphics[width=0.5\columnwidth]{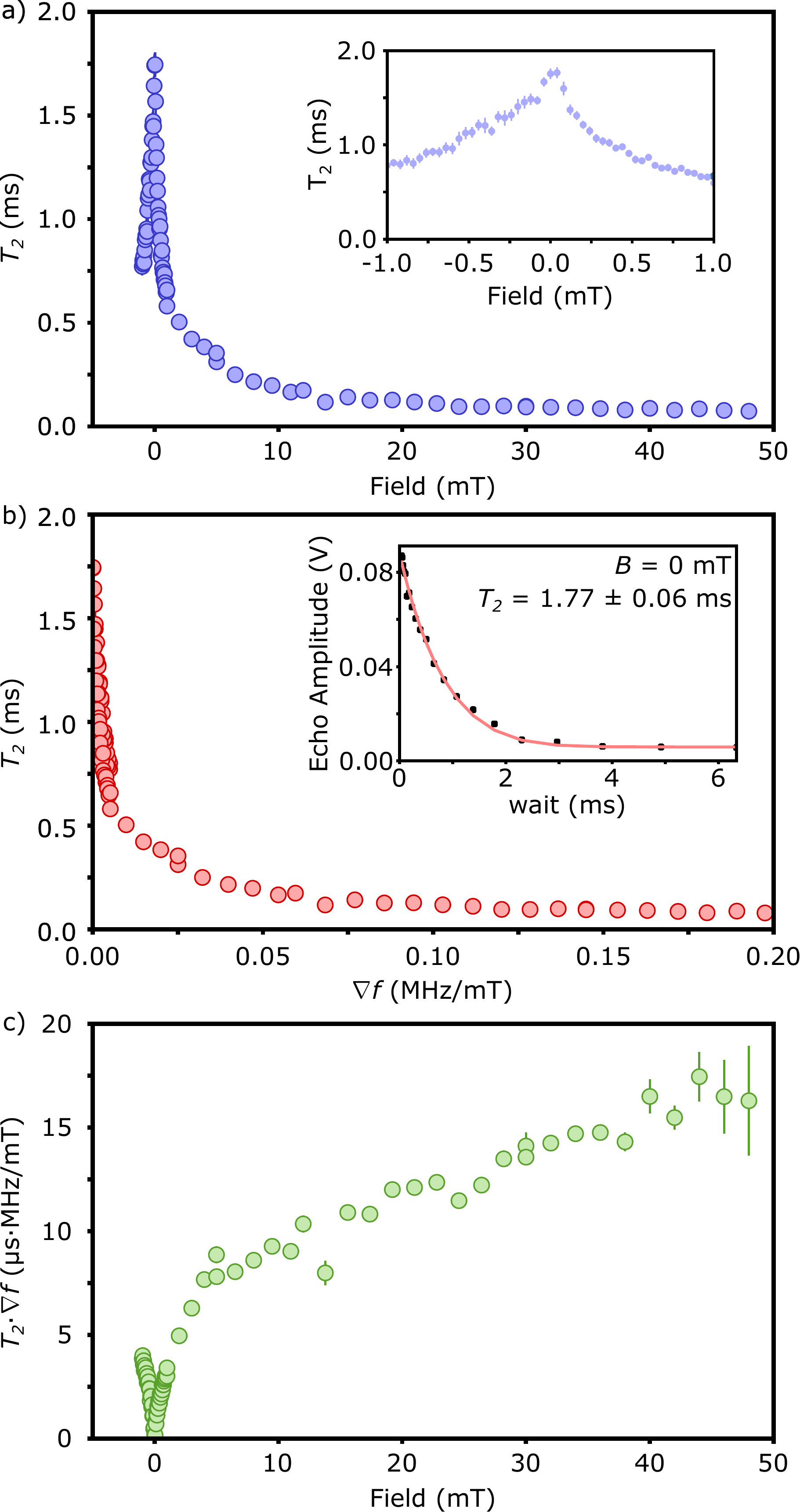}
    \caption{Summary of $T_{2}$ data around zero field in \yb{nat}. The coherence time reaches a maximum at \SI{0}{\milli\tesla} (a), this corresponds to the minimum in effective g-factor - $\nabla_{B} f$ (b). c) The product of $\nabla_{B} f$ and $T_{2}$, as the sensitivity to spectral diffusion is proportional to $\frac{1}{\nabla_{B}}f$ the product should be constant for constant spectral diffusion. We observe that the product decreases at \SI{0}{\milli\tesla} which means spectral diffusion is increasing faster than the \yb{171} spin can become insensitive to it. This increase in spectral diffusion is caused by unpolarised \yb{} isotopes, especially those with zero nuclear spin.}
    \label{fig:zefoz T2s}
\end{figure}

FIG. \ref{fig:zefoz T2s} gives a summary of the measured coherence times along this line of low $\nabla_{B}f$. We observe a substantial increase in $T_{2}$ as \SI{0}{\milli\tesla} is approached (a) which corresponds to a decrease in the effective g-factor (b). To understand why the $T_{2}$ at zero field is much less than that measured at high field we calculate the product of $T_{2}$ and $\nabla_{B} f$. Spectral diffusion is proportional to $\frac{1}{\nabla_{B} f}$ so the product $T_{2}\cdot \nabla_{B} f$ is constant for constant environmental spectral diffusion. As we approach zero field the rate of spectral diffusion increases as environmental spins become completely unpolarised - particularly \yb{} isotopes with zero nuclear spin. We observe that $T_{2}\cdot \nabla_{B} f$ decreases substantially as \SI{0}{\milli\tesla} is approached, this means the rate of spectral diffusion increases faster than the \yb{171} spin becomes insensitive to it. This limits the usefulness of the zero field clock transition of \yb{171} in \yb{nat} doped systems.

\section{ESEEM in $^{171}$Yb:Y$_{2}$SiO$_{5}$}
\begin{figure}
    \centering
    \includegraphics[width=0.5\columnwidth]{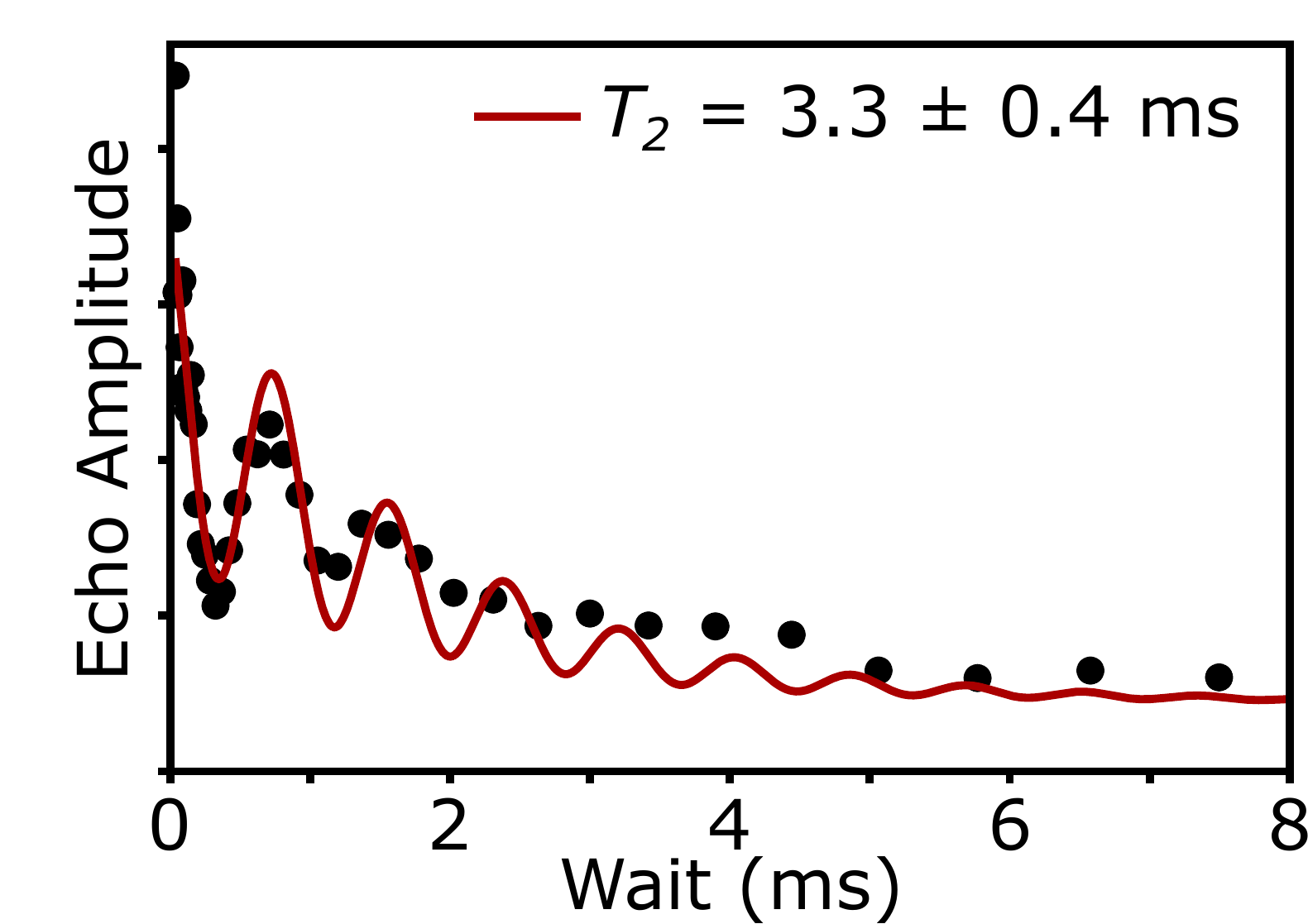}
    \caption{ESEEM in \yb{171}:\yso, the super-hyperfine interaction between \yb{171} electron spins and \y{89} nuclear spins modulates the echo amplitude in a two-pulse echo sequence. By fitting a damped cosine with frequency equal to the Larmor frequency of \y{89} a $T_{2}$ time can be extracted.}
    \label{fig:ESEEM}
\end{figure}

Electron spin echo envelope modulation (ESEEM) was observed at low fields ($<$\SI{1}{\milli\tesla}) in the isotopically pure \yb{171} sample (Fig.~\ref{fig:ESEEM}). This occurs when the super-hyperfine interaction between \yb{171} electron spins and \y{89} nuclear spins modulates the echo amplitude in a two-pulse (Hahn) echo sequence. This results in a perceived decrease in the coherence time, however a damped cosine can be fitted where the frequency is equal to the Larmor frequency of \y{89}. However, at \SI{0}{\milli\tesla}, this frequency goes to zero and the period of oscillation goes to infinity. This means the quantum state is completely transferred to the \y{89} nuclear spin at zero field and is irretrievable, this prevents the zero field clock transition being used as a quantum memory without suppression of the super-hyperfine interaction.




\newpage
\bibliographystyle{unsrt}
\bibliography{bibsi}